\documentclass[12pt]{article}
\usepackage{epsfig}
\usepackage{amssymb}
\def\al{\alpha}

\def\la{\lambda}
\def\th{\theta}

\def\si{\sigma}

\newcommand{\beq}{\begin{equation}}
\newcommand{\eeq}{\end{equation}}
\newcommand{\bea}{\begin{eqnarray*}}
\newcommand{\eea}{\end{eqnarray*}}
\newcommand{\beqa}{\begin{eqnarray}}
\newcommand{\eeqa}{\end{eqnarray}}

\newcommand{\txi}{{\tilde\xi}}

\newcommand{\mbar}{{\overline m}}
\newcommand{\psib}{{\overline \psi}}
\newcommand{\limi}[1]{\raisebox{-0.23cm}{~\shortstack{ $\mbox{lim
}$ \\
${\vspace{-0.2cm} _{#1}}$}}}

\begin{document}
\newfont{\elevenmib}{cmmib10 scaled\magstep1}%

\newcommand{\preprint}{
            \begin{flushleft}
   \elevenmib Yukawa\, Institute\, Kyoto\\
            \end{flushleft}\vspace{-1.3cm}
            \begin{flushright}\normalsize  \sf
            YITP-02-01\\
           {\tt hep-th/0203085} \\ March 2002
            \end{flushright}}
\newcommand{\Title}[1]{{\baselineskip=26pt \begin{center}
            \Large   \bf #1 \\ \ \\ \end{center}}}
\hspace*{2.13cm}%
\hspace*{1cm}%
\newcommand{\Author}{\begin{center}\large
           Pascal Baseilhac\footnote{
pascal@yukawa.kyoto-u.ac.jp}
\end{center}}
\newcommand{\Address}{{\baselineskip=18pt \begin{center}
           \it Yukawa Institute for Theoretical Physics\\
     Kyoto University, Kyoto 606-8502, Japan
      \end{center}}}
\baselineskip=13pt

\preprint
\bigskip

\Title{Liouville field theory coupled to a critical Ising model:
Non-perturbative analysis, duality and applications} \Author

\Address

\vskip 1.1cm

\centerline{\bf Abstract}

\vspace{0.2cm} Two different kinds of interactions between a
${\mathbb Z }_{n}$-parafermionic and a Liouville field theory are
considered. For generic values of $n$, the effective central
charges describing the UV behavior of both models  are calculated
in the Neveu-Schwarz sector. For $n=2$ exact vacuum expectation
values of primary fields of the Liouville field theory, as well as
the first descendent fields are proposed. For $n=1$, known results
for Sinh-Gordon and Bullough-Dodd models are recovered whereas for
$n=2$, exact results for these two integrable coupled
Ising-Liouville models are shown to exchange under a weak-strong
coupling duality relation. In particular, exact relations between
the parameters in the actions and the mass of the particles are
obtained. At specific imaginary values of the coupling and
 $n=2$, we use previous results to obtain exact information about:
  (a) Integrable coupled models like Ising-${\cal M}_{p/p'}$, homogeneous sine-Gordon model
  $SU(3)_2$ or the Ising-XY model, (b) Neveu-Schwarz sector of the $\Phi_{13}$
integrable perturbation of $N=1$ supersymmetric minimal models.
Several non-perturbative checks are done, which support the exact
results.\\
{\small PACS: 10.10.-z; 11.10.Kk; 11.25.Hf; 64.60.Fr}
\vskip 0.8cm


\vskip -0.6cm

{{\small  {\it \bf Keywords}: Massive integrable field theory;
Duality; Coupled models; Homogeneous sine-Gordon;}}\\
\vspace{-0.45cm}

{{\small Vacuum expectation values; $N=1$ Superconformal minimal
models}}
\vskip 1cm

\section{Introduction}
Duality plays an important role in the analysis of quantum field
theory (QFT) and statistical physics. This property allows to
study the behavior of observable in the strong coupling region of
one model in terms of the ones in the weak coupling region of the
other (dual) model. In different region of the coupling constant,
it is then possible to use perturbative and semiclassical methods.
For instance, in four dimensions the electro-magnetic duality
conjectured in \cite{Oli} and developed in \cite{Sei} is the main
ingredient in studying the spectrum and phase structure in $N=2$
supersymmetric Yang-Mills theory.

Almost all two-dimensional relativistic theories can be understood
as conformal field theories  (CFT)s  - describing a fixed point of
the theory -  perturbed by a relevant operator. If the
perturbation preserves integrability, then the analysis remains
exact {\it beyond} this point. In such cases, the non-perturbative
analysis essentially simplifies. Besides its Lagrangian
formulation, the QFT also possess an unambiguous definition in
terms of factorized scattering theory. These data permit one to
use non-perturbative methods for the analysis of integrable QFTs
and make it possible, in some cases, to justify the existence of
two (dual) Lagrangian representations of the theory.

Among two-dimensional dual models, one finds the well-known
sine-Gordon/massive Thirring QFTs \cite{Col}. Another interesting
example of duality is the weak-strong coupling duality flow from
the (affine) Toda theories (A)TFTs  based on the (affine) Lie
algebra ${\cal G}$ to the theory based on the dual (affine) Lie
algebra $\tilde{\cal G}$ \cite{atft}. More generally, integrable
deformations of ATFTs also provide series of dual models
\cite{del,supa,dual} possessing many applications. For instance,
the scattering data ($S$-matrix) for the ones based on Lie
superalgebras with massive excitations only have been considered
in detail \cite{del,supa}. These models correspond  to ATFTs (for
the purely bosonic part) coupled with one or two Majorana
fermions. This was done either for models based on
$A^{(2)}(0,2r-1)$ and $C^{(2)}(r+1)$ (whose bosonic root subsystem
is simply laced) but also for $B^{(1)}(0,r)$ and $A^{(4)}(0,2r)$
(non-simply laced subsystem corresponding respectively to $BC_r$
and $C_r^{(1)}$ roots). In particular, these cases can be obtained
as reductions of more general deformed ATFTs \cite{dual}.
Furthermore, they can be seen as integrable perturbations of CFTs
with extended symmetry based on $WB(0,r)$ (or also called
fermionic $WB_{r}$) algebras.

Let us now consider some of the simplest cases above, i.e.
corresponding to a rank $r=1$. Then the models $C^{(2)}(2)$,
$B^{(1)}(0,1)$ and $A^{(4)}(0,2)$ can be understood as integrable
perturbations of $WB(0,1)$ CFT, i.e. the $N=1$ superLiouville (SL)
field theory. Integrability can be shown either by using the
so-called singular vector analysis \cite{watts} or by constructing
explicitly local conserved quantities (see for instance
\cite{moda} for the two latter cases). In the first model
$C^{(2)}(2)$ the perturbation preserves supersymmetry and it
corresponds to the well-known supersinh-Gordon model. In the two
other cases, which will be the subjects of the present analysis,
supersymmetry is broken.

If the theory contains only massive particles (like those
presented above), the $S$-matrix data exhibit information about
the long distance (IR) property of the theory. On the other hand,
the knowledge of the CFT data and the relevant operator associated
with the perturbation also define completely the theory. Once
identified, the CFT contains all information about the UV behavior
of the theory. It is consequently important to connect these two
kinds of data in order to understand better the structure of
integrable QFT. When the basic CFT admits a representation of the
primary operators in terms of vertex operators the CFT data
contain the so-called ``reflection amplitudes'', which relate
different vertex operators possessing the same quantum numbers.
Considering the system on a circle of size $R$, it has been shown
\cite{ZamZam} that reflection amplitudes play a crucial role in
the determination of the effective central charge $c_{eff}(R)$ in
the UV domain ($R\rightarrow 0$) of the QFT. On the other hand,
the effective central charge $c_{eff}(R)$ can be calculated
independently from the $S$-matrix data using the TBA method. At
small $R$, asymptotic can be compared with the one following from
CFT data, if one knows explicitly the exact relation between the
mass of the particles (IR data) and the (UV) parameter in the
Lagrangian. The agreement of both approaches can be considered as
a non trivial test for the $S$-matrix elements. On the other hand,
reflection amplitudes are a key tool \cite{onep} in the
calculations of exact vacuum expectation values (VEVs) in
integrable QFTs. It has also been shown that perturbative analysis
and semiclassical methods support these conjectures. Both types of
non-perturbative analysis have been performed in many cases
\cite{ZamZam,onep,ahn,norma,atfttba1,atfttba2,basfat}. Beside the
confirmation of the $S$-matrix and calculation of exact
quantities, they also provide a useful tool to check duality.

Perturbed CFTs have also attracted much attention since they can
describe many physical systems in the vicinity of a critical
point. Several models connected with either ATFTs\,\footnote{For
instance, the analytic continuation of $A_{n-1}$ ATFT provides
exact results for leading thermal perturbation of
$Z_n$-parafermionic CFTs \cite{norma}. Also, integrable
perturbations of minimal models have been considered \cite{onep}.}
or integrable deformations of ATFTs have received particular
attention in the recent years, for a large class of strongly
interacting solid state physics problems.

In a recent paper \cite{coup1} we considered integrable coupled
identical minimal models for which the interaction preserves
integrability. These kind of models were introduced in
\cite{zN,vays} for which the scattering properties were
considered. In \cite{coup1}, exact results like VEVs and
mass-parameter relations are obtained using exact results for
$C^{(1)}_2$ ATFTs \cite{atfttba2}.

In this paper, we consider in detail the ATFTs based on super Lie
algebras $B^{(1)}(0,1)$ and $A^{(4)}(0,2)$ which correspond to a
critical Ising model coupled with  a Liouville field theory. Using
non-perturbative analysis based on reflection amplitudes, in the
UV limit of both models the effective central charges are
calculated in the Neveu-Schwarz (NS) sector (subsect. 2.1). Exact
VEVs of primary fields  which belong to the Liouville field theory
and their first descendants are proposed (sect. 3). Observables of
both models are shown to interchange with the flow of the coupling
$b$. In particular they coincide at the self-dual point $b^2=1$.
Duality relations between parameters in both Lagrangian are
proposed as well as exact relations with the mass of the
particles.

In section 4, taking analytic continuation of the coupling
$b\rightarrow i\beta$ and fixing its value, several kind of
coupled minimal models can be obtained. Namely, a critical Ising
model and minimal model ${\cal M}_{p/p'}$ interacting through \
$\epsilon(x) \Phi_{12}(x)$ or\ $\epsilon(x) \Phi_{21}(x)$\ where
$\epsilon(x)$ denotes the energy operator of the critical Ising
model and $\Phi_{12}(x)$ and $\Phi_{21}(x)$ are primary operators
of the second model. For instance, together with the results of
\cite{coup1} this completes the analysis for Ising-Ising coupled
models, as energy-spin interactions are considered here. Among
other models,
 exact results are obtained for the Ising-3-state Potts models coupled by energy-energy which
appear in the phase diagram of ${\mathbb Z}_6$ spin models
\cite{doreyt}. Furthermore, we will relate the critical and
tricritical Ising models coupled through this type of interaction
to a special case of the homogeneous sine-Gordon model (HSG)
$SU(3)_2$ \cite{mir0} (see \cite{hsg} and references therein).
This model which can be understood as an integrable perturbation
of the WZNW-coset CFT $SU(3)_2/(U(1))^2$ have recently attracted
attention \cite{mir0,mir,mir2,hsg}. It possesses applications in
one-dimensional quantum spin systems: such perturbations appear
directly in the CFT study of a three-leg spin ladder which
consists of three spin-$1/2$ Heisenberg chains weakly coupled by
on-rung interaction \cite{Phil}.

However, there are also other coupled models which, although
studied numerically, have not yet been studied analytically in
great detail. For instance, the coupled Ising-XY model has
attracted great attention as it is expected to describe the
critical behavior of a large class of two dimensional classical XY
models having the particularity to exhibit both
 discrete ${\mathbb Z}_2$ and continuous $U(1)$ degeneracy in
their ground state. Among these models, one finds the fully
frustrated XY model (FFXY) \cite{be,gra}, $J_1$-$J_2$ XY model
\cite{pascal}, triangular lattice frustrated XY model \cite{dhl},
helical XY model \cite{vil}, the Coulomb gas system of
half-integer charges \cite{Thij},... In particular, the FFXY model
can be physically realized as a Josephson-junction array of large
capacitance in a perpendicular magnetic field corresponding to a
half-flux quantum per plaquette \cite{jos}. For imaginary values
of the coupling, both models considered here\,\footnote{The model
associated with $C^{(2)}(2)$, which corresponds to the $N=1$
supersymmetric sine-Gordon model, has been suggested as a
candidate for the coupled Ising-XY model by Foda \cite{foda}.
Exact results for this case follow from \cite{basfat} so we will
not discuss this model here.} possess such ${\mathbb Z}_2\times
U(1)$ symmetry, so that we apply previous results in section 5.

The models $B^{(1)}(0,1)$ and $A^{(4)}(0,2)$ describe integrable
perturbations of $WB(0,1)$ superconformal minimal models, i.e.
superLiouville theory. At specific values of the coupling, one can
then obtain VEVs and exact mass-UV parameter relation in $N=1$
superconformal minimal models perturbed by Neveu-Schwarz operators
$\Phi^{(NS)}_{13}$, $\Phi^{(NS)}_{31}$ or $\Phi^{(NS)}_{15}$. For
simplicity, only the first case is considered in detail in section
6. Some checks are done which support the results.

Finally, some remarks about the Ramond (R) sector of both models
follow in the last section.

\section{Liouville field theory coupled to ${\mathbb Z }_{n}$-parafermions}
Instead of studying coupled Ising-Liouville models we rather
prefer to consider two different kinds of interactions between a
${\mathbb Z }_{n}$-parafermionic \cite{fatzam} and a Liouville
field theory. Although probably not integrable for generic values
of $n\neq 1,2$, it provides a useful way to check the consistency
of the expressions for $n=1,2$.  Consequently, we consider two
QFTs which admit Lagrangian representations described in terms of
a parafermionic field theory interacting with a Liouville-like
term
\beqa
 {\cal A}_{\tau=1}^{(n)}&=&{\cal A}^{(n)}_0 + \int d^2x\Big[
\frac{1}{8\pi}(\partial_\nu\varphi)^2 -\kappa\psi{\overline
\psi}e^{-b\varphi} + \mu e^{b\varphi}\Big]\ ,\label{a1}\\
{\cal A}_{\tau=2}^{(n)}&=&{\cal A}^{(n)}_0 + \int d^2x\Big[
\frac{1}{8\pi}(\partial_\nu\varphi)^2
-{\tilde\kappa}\psi{\overline \psi}e^{-b\varphi} + {\tilde\mu}
e^{2b\varphi}\Big]\ .\label{a2} \eeqa
Here ${\cal A}^{(n)}_0$ denotes the ${\mathbb{Z}}_n$ parafermionic
CFT with central charge $c=2-\frac{6}{n+2}$ and the fields
$\psi(\psib)$ are the holomorphic (antiholomorphic) parafermionic
currents with spin $s=1-\frac{1}{n}$ (${\overline s}=-s$). For
$n=1$, we have $\psi(\psib)={\mathbb I}$ in (\ref{a1}), (\ref{a2})
: the Lagrangians above coincide with the (integrable) well-known
Sinh-Gordon (ShG) and Bullough-Dodd (BD) model, respectively. For
$n=2$, the parafermionic current $\psi$ is a Majorana fermion and
to cancel fermion divergencies we have to add a counterterm
${\tilde\mu}e^{-2b\varphi}$ in both cases (\ref{a1}), (\ref{a2}).
The resulting Toda-like part in the QFTs (\ref{a1}), (\ref{a2})
then becomes respectively a BD or a ShG model. It is known that
these QFTs are integrable for $n=2$ for which the factorized
scattering theory has been studied in detail in \cite{del}.

Like for most of two dimensional field theories, conformal
perturbation theory (CPT) can be used to study these models.
Indeed, QFT with action (\ref{a1}) (or (\ref{a2}) with the
substitution $\kappa\rightarrow{\tilde\kappa}$) can be seen as two
different deformations of the same following CFT
\beqa {\cal A}_{CFT}^{(n)}={\cal A}^{(n)}_0 + \int d^2x\Big[
\frac{1}{8\pi}(\partial_\nu\varphi)^2 -\kappa\psi{\overline
\psi}e^{-b\varphi}\Big]\ \label{CFT1} \eeqa
which, for $n=1,2$, coincides with the Liouville and $N=1$
supersymmetric Liouville models (adding the counterterm),
respectively. The stress-energy tensor
\beqa
T^{(n)}(z)=T^{(n)}_P(z)-\frac{1}{2}(\partial\varphi)^2-Q'\partial^2\varphi\qquad
\mbox{with}\qquad Q'=\frac{b}{2}+\frac{1}{nb} \label{Tn}\eeqa
and $T^{(n)}_P$ associated with the parafermionic CFT, ensures the
local conformal invariance of (\ref{CFT1}). Here, we denote \
$\partial = \frac{1}{2}(\partial_x-i\partial_y)$ and the fields
are normalized such that
\beqa \langle\varphi(x)\varphi(y)\rangle_{Gaussian}=-\ln|x-y|^2\
.\eeqa
For instance, we have the conformal dimensions
\beqa
\Delta(\psi{\overline\psi}e^{a\varphi})=1-\frac{1}{n}-\frac{a(2Q'+a)}{2}\
. \label{dimpert}\eeqa
Besides the conformal symmetry, for general values of $n$ the
${\mathbb Z}_n$ parafermionic CFT possesses an additional symmetry
generated by the parafermionic current $\psi(\psib)$
\cite{fatzam}. The basic fields in this CFT are the order
parameters $\sigma_j$, $j=0,1,...,n-1$ with conformal dimensions \
$\delta_j=j(n-j)/(2n(n+2))$. The operator algebra also contains
${\mathbb Z}_n$ neutral fields $\epsilon^{(j)}$, $j=1,...\leq n/2$
with conformal dimensions \ $D_j=j(j+1)/(n+2)$. All fields in this
CFT can be obtained from the field $\sigma_j$ by application of
the generators of the parafermionic symmetry \cite{fatzam}. It is
thus natural to introduce as the basic operators the local fields
$\sigma_je^{a\varphi}$. However, here we will essentially focus on
the Liouville part, i.e. we will take $j=0$.

 The exponential fields $V_a=e^{a\varphi}$ are spinless conformal primary
fields of the CFT (\ref{CFT1}), with conformal dimensions
$-a(2Q'+a)/2$. It can be shown \cite{basfat} that the fields $V_a$
and $V_{-2Q'-a}$ are reflection images of  each other, i.e. they
are related by the linear transformation
\beqa e^{a\varphi}=R_b^{(n)}(a)\ e^{-(2Q'+a)\varphi}\
.\label{foncn}\eeqa
For $n=1$, the reflection amplitude $R_b^{(n)}(a)$ reduces to the
so-called ``Liouville reflection amplitude'' proposed in
\cite{ZamZam}. For $n=2$, it corresponds to the $N=1$
superLiouville reflection amplitude in the NS sector proposed in
\cite{StanRash}. For general values $n$, the (NS for $n=2$)
reflection amplitude $R_b^{(n)}(a)$ associated with action
(\ref{CFT1}) writes \cite{basfat}
\beqa
R^{(n)}_{b}(a)=-\left[\frac{\pi\kappa\gamma(Q'b)}{n\big(\frac{n^2b^4}{4}\big)^{1/n}}\right]^{\frac{2(a+Q')}{b}}
\frac{\Gamma(1-(a+Q')b)\Gamma(1-2(a+Q')/nb)}{\Gamma(1+(a+Q')b)\Gamma(1+2(a+Q')/nb)}\
. \label{refn} \eeqa
Here we denote $\gamma(x)=\Gamma(x)/\Gamma(1-x)$ as usual.

\subsection{Scaling functions for generic values of $n$}
Consider now the CFT (\ref{CFT1}) on an infinite cylinder of
circumference $2\pi$ with the cartesian coordinates $x_1$, $x_2$
where $x_2$ along the cylinder is defined as the imaginary time
and $x_1\sim x_1+2\pi$ is the space coordinate. For simplicity, we
will here focus on the sector in which only bosonic zero-modes
appear. In particular,  for $n=2$ it corresponds to the NS sector.
Then, the reflection amplitude (\ref{refn}) provides
 non-perturbative information in the study of quantum mechanical problem for
bosonic zero
modes
\beqa \varphi_0=\int_{0}^{2\pi}\varphi(x)\frac{dx_1}{2\pi}\eeqa
of the fields $\varphi(x)$.  In the semi-classical limit
$b\rightarrow 0$, where one can neglect the oscillator modes of
$\varphi(x)$, the Schr\"{o}dinger equation governing the zero-mode
dynamics writes\,\footnote{For $n=2$, one has for instance
\cite{ahn} \ ${\cal H}^{(2)}_0 \equiv
-\frac{1}{8}-\big(\frac{\partial}{\partial\varphi_0}\big)^2
 + 2\pi{\tilde\mu} e^{-2b\varphi_0}$}
\beqa {\cal H}^{(n)}_0\Psi_P(\varphi_0)=E_0\Psi_P(\varphi_0)
\qquad \mbox{with}\qquad E_0=-\frac{n+1}{24}+P^2\ \label{E0} \eeqa
the ground state energy where the momentum $P$ is a real vector.
The full quantum effect can be implemented simply by introducing
the exact reflection amplitudes which take into account also non
zero-mode contributions \cite{ZamZam}.

The wave function in the asymptotic region can be obtained using
the same arguments as the ones already applied for other models.
Namely, the exponential term in the Hamiltonian is considered as a
potential wall. An incident plane wave with momentum $P$ is then
reflected to the plane wave with $-P$. The phase change
corresponding to this process is associated with the reflection
amplitude given above. Consequently, the wave function
$\Psi_P(\varphi_0)$ is simply written as a superposition of two
plane waves:
\beqa \Psi_P(\varphi_0)\simeq e^{iP\varphi_0} +
S_b^{(n)}(P)e^{-iP\varphi_0}\qquad \mbox{at}\qquad
\varphi_0\rightarrow +\infty\ . \eeqa
with $S_b^{(n)}(P)=R_b^{(n)}(-iP-Q')$. Using the approach proposed
in \cite{ZamZam}, we can now obtain the scaling functions in the
UV region of the QFTs (\ref{a1}) or (\ref{a2}) defined on a
cylinder with circumference $R\rightarrow 0$.

Let us first consider the QFT (\ref{a1}). The additional term $\mu
e^{b\varphi}$ in its action compared to the CFT introduces a new
potential wall, which implies the quantization of the momentum $P$
in the wave function. It depends on the size of the enclosed
region, which is proportional to $\ln(1/R)$. As we will see later,
 this quantized momentum $P(R)$ defines  the scaling function
$c_{eff}(R)$ in the UV region using eq. (\ref{E0}).

In action (\ref{a1}) the dimensions of the parameters are
$Dim[\kappa]=\frac{2}{n}+b^2$ and $Dim[\mu]=2+b^2$. It is now
convenient to rescale back the size of the system from $R$ to
$2\pi$. The action (\ref{a1}) becomes
\beqa {\cal A}_{\tau=1}^{(n)}={\cal A}^{(n)}_0 + \int d^2x\Big[
\frac{1}{8\pi}(\partial_\nu\varphi)^2
-\kappa\big(\frac{R}{2\pi}\big)^{\frac{2}{n}+b^2}\psi{\overline
\psi}e^{-b\varphi}+\mu\big(\frac{R}{2\pi}\big)^{2+b^2}e^{b\varphi}\Big]\nonumber
\ . \eeqa
Following previous analysis (see \cite{ZamZam,ahn} for instance)
and due to the form of the perturbing term in (\ref{a1}) we obtain
the following quantization condition for the momentum $P$:
\beqa
\big(\frac{R}{2\pi}\big)^{-2iP(b+1/b)H_n}S_b^{(1)}(P)|_{\kappa\rightarrow
-\mu}S_b^{(n)}(P)=1\ , \label{quantcond} \eeqa
where $S_b^{(1)}(P)$ is nothing but the so-called ``Liouville
reflection amplitude'' proposed in \cite{ZamZam}. For further
convenience, here we introduced the ``deformed'' Coxeter number
\cite{Cor2}:
\beqa H_n=\frac{2(n+1)}{n}(1-B)+2B\qquad \mbox{with}\qquad
B=\frac{b^2}{1+b^2}\ .
 \eeqa
For the lowest energy state, in terms of the reflection phases,
eq. (\ref{quantcond}) reduces to
\beqa LP=2\pi-\delta^{(1)}_{b}(P)-\delta^{(n)}_{b}(P)
\eeqa\label{eqpert}
with
\beqa
L=-\frac{2}{b}H_n(1+b^2)\ln\big(\frac{R}{2\pi}\big)-\frac{1}{b}
\ln\Big[\pi\mu\gamma(b^2/2)\frac{\pi\kappa}{n}\frac{\gamma(1/n+b^2/2)}{(\frac{n^2b^4}{4})^{1/n}}\Big]^2\nonumber
\eeqa
and we used the convenient notation
\beqa
\delta^{(n)}_{b}(P)=-i\ln\Big[\frac{\Gamma(1+iPb)\Gamma(1+i2P/nb)}{\Gamma(1-iPb)\Gamma(1-i2P/nb)}\Big]\
. \nonumber\eeqa
In the UV region $R\rightarrow 0$ we can solve eq. (14)
perturbatively by expanding the reflection phases in powers of
$P$. We obtain
\beqa \ell
P=2\pi-\big(\delta^{(1)}_{b,3}+\delta^{(n)}_{b,3}\big)P^3
-\big(\delta^{(1)}_{b,5}+\delta^{(n)}_{b,5}\big)P^5+...\label{expand}
\eeqa
where we define
\beqa \delta^{(n)}_{b,1}=-2\gamma_E\big(b+\frac{2}{nb}\big)\ ,
\qquad
\delta^{(n)}_{b,s}=(-)^{\frac{s-3}{2}}\frac{2}{s}\zeta(s)\big(b^s+\big(\frac{2}{nb}\big)^s\big)\nonumber
\eeqa
and introducing the Euler constant $\gamma_E$,
\beqa \ell\equiv L-L_0=L-2\gamma_E\frac{H_n(1+b^2)}{b}\
.\label{ell} \eeqa
The ground state energy of the system on the circle of size $R$ is
 given by $E(R)=-\pi c_{eff}(R)/6R$ \ with the effective central charge \
$c_{eff}(R)=(n+1)/2-12P^2$. Here, $P$ is the solution of the above
quantization condition (14), and perturbatively (\ref{expand}) in
powers of $1/\ell$. After some calculations, we find that the UV
behavior of the QFT (\ref{a1}) - in the NS sector for $n=2$ - is
characterized by the scaling function
\beqa
c^{(n)}_{eff}(R)&=&\frac{n+1}{2}-12\Big[\big(\frac{2\pi}{\ell}\big)^2
-\frac{2}{3}\frac{\zeta(3)\big(2b^3+8(n^3+1)/n^3b^3\big)}{\pi}\big(\frac{2\pi}{\ell}\big)^5\nonumber\\
&&\ \ \ \ \ \ \  +\
\frac{2}{5}\frac{\zeta(5)\big(2b^5+32(n^5+1)/n^5b^5\big)}{\pi}\big(\frac{2\pi}{\ell}\big)^7
+ {\cal O}\big(\big(\frac{2\pi}{\ell}\big)^8\big)\Big] \
.\label{ceff}\eeqa
In particular, for $n=1$ this result agrees perfectly with the
sinh-Gordon one \cite{ZamZam}.

 The same analysis can be performed along the same line for a different kind of perturbation,
 i.e. for instance in QFT (\ref{a2}). In this case, the dimension of the
parameters in the action are $Dim[\tilde\kappa]=\frac{2}{n}+b^2$
and $Dim[\tilde\mu]=2+4b^2$. The quantization condition is now
\beqa \big(\frac{R}{2\pi}\big)^{-2iP(b+1/b){\tilde
H}_n}S_{2b}^{(1)}(P)|_{\kappa\rightarrow
-{\tilde\mu}}S_b^{(n)}(P)=1 \ ,\label{quantcondt}\eeqa
where we define the ``deformed'' Coxeter number ${\tilde
H}_n=\frac{n+2}{n}(1-B)+3B$. As before, eq. (\ref{quantcondt}) can
be written in terms of the reflection phases as ${\tilde
L}P=2\pi-\delta^{(1)}_{2b}(P)-\delta^{(n)}_{b}(P)$ with
\beqa {\tilde L}=-\frac{2}{b}{\tilde
H}_n(1+b^2)\ln\big(\frac{R}{2\pi}\big)-\frac{1}{b}
\ln\Big[\big[\pi{\tilde
\mu}\gamma(2b^2)\big]\big[\frac{\pi{\tilde\kappa}}{n}\frac{\gamma(1/n+b^2/2)}{(\frac{n^2b^4}{4})^{1/n}}\big]^2\Big]\
. \eeqa
Consequently, it is straightforward to show that the scaling
function for the UV behavior of the QFT (\ref{a2}) - in the NS
sector for $n=2$ - is given by the following expansion
\beqa {\tilde
c}^{(n)}_{eff}(R)&=&\frac{n+1}{2}-12\Big[\big(\frac{2\pi}{{\tilde\ell}}\big)^2
-\frac{2}{3}\frac{\zeta(3)\big(9b^3+(8+n^3)/n^3b^3\big)}{\pi}\big(\frac{2\pi}{{\tilde\ell}}\big)^5\nonumber\\
&&\ \ \ \ \ \ \  +\
\frac{2}{5}\frac{\zeta(5)\big(33b^5+(32+n^5)/n^5b^5\big)}{\pi}\big(\frac{2\pi}{{\tilde\ell}}\big)^7
+ {\cal O}\big(\big(\frac{2\pi}{{\tilde\ell}}\big)^8\big)\Big]\
\label{cefft} \eeqa
with ${\tilde \ell}$ similar to $\ell$ in (\ref{ell}) but with the
replacement $H_n\rightarrow{\tilde H}_n$. It can be checked that
for $n=1$, this result coincides exactly with the Bullough-Dodd
\cite{ahn} scaling function, as expected.

Notice that for the specific value of the coupling constant
 $b^2=1$, both scaling functions take the same form
for any values of $n$.

\subsection{Thermodynamic Bethe ansatz for $n=2$ and duality}
For specific values of $n$, the effective central charge
calculated above from the CFT data (reflection amplitudes) can be
compared with the same function determined from the numerical
solution of the TBA equations for the QFT (\ref{a1}) and
(\ref{a2}). For $n=1$, this analysis has been done in
\cite{ZamZam} for the ShG and in \cite{ahn} for the BD cases.
Then, let us consider the case $n=2$ in the QFTs (\ref{a1}) and
(\ref{a2}). As was conjectured in \cite{del} (even for higher rank
$r > 1$ of the affine Toda part) these QFTs possess a weak-strong
coupling duality: the analysis of the factorized scattering theory
shows that there exists a QFT which possesses two (dual)
perturbative regimes associated with action (\ref{a1}) and
(\ref{a2}), respectively. An intermediate mass spectrum
(consisting of two particles for the rank $r=1$) was proposed:
\beqa M_{\psi}=\mbar \qquad \mbox{and}\qquad
M_1=2\mbar\sin(\pi/H_2)\ . \label{spect}\eeqa
This mass spectrum flows from the one of the QFT (\ref{a1}) to the
one associated with the QFT (\ref{a2}) while $b$ increases.

Using the corresponding (diagonal) FST proposed in \cite{del} - we
report the reader to this work for details - one writes the TBA
equations. Namely,
\beqa c_{eff}^{(TBA)}(R)=\frac{3R}{\pi^2}\int\cosh\theta
\Big[M_\psi\ln(1+e^{-\epsilon_\psi(\theta,R)})+M_1\ln(1+e^{-\epsilon_1(\theta,R)})\Big]d\theta\label{ceffTBA}\eeqa
where the functions $\epsilon_{\psi}(\theta,R)$,
$\epsilon_{1}(\theta,R)$ satisfy the system of 2 coupled integral
equations
\beqa M_iR\cosh\theta =
\epsilon_{i}(\theta,R)+\sum_{j\in\{1,\psi\}}\int\varphi_{ij}(\th-\th')\ln(1+e^{-\epsilon_j(\th',R)})
\frac{d\th'}{2\pi}\qquad \mbox{for}\qquad i\in\{1,\psi\}\nonumber
\eeqa
with the kernels $\varphi_{ij}$ defined as the logarithmic
derivatives of the $S$-matrix elements obtained in \cite{del}.
However, the function $E^{(TBA)}(R)$ defined from the TBA
equations differs from the ground state energy $E(R)$ of the
system on the circle of size $R$ by the bulk term
$E^{(TBA)}(R)=E(R)-f^{(2)}_{\tau}R$, where $f^{(2)}_\tau$ is a
specific bulk free energy of the QFT. To compare the same
functions we should then substract this term from the function
$E(R)$ defined by $c^{(2)}_{eff}(R)$, i.e. for instance in case
(\ref{a1}) we have
\beqa c_{eff}^{(TBA)}(R)=c_{eff}^{(2)}(R) +
\frac{6R^2}{\pi}f^{(2)}_{\tau=1}\  \eeqa\label{receff}
and similarly for case (\ref{a2}). Notice that the contribution of
bulk term $f^{(2)}_{\tau}$ becomes quite essential at $R\sim {\cal
O}(1)$. In many examples of known QFTs, this quantity can be
calculated explicitly using Bethe Ansatz (BA) method (see for
instance \cite{zamba,fatba}). Here, we propose the following
expression for the specific bulk free energy in the QFT (\ref{a1})
(the same quantity for the QFT (\ref{a2}) follows using the
duality $B\rightarrow 1-B$ and $H_2\rightarrow {\tilde H}_2$):
\beqa
f^{(2)}_{\tau=1}=\frac{\mbar^2}{8}\frac{\sin(\pi/H_2)}{\sin(\pi
B/H_2)\sin(\pi(1-B)/H_2)}\label{f}\ . \eeqa
Finally, to compare the expansion coming from the CFT data and the
one from the numerical analysis of TBA eqations, we need the exact
relations between the UV parameters in the actions (\ref{a1}) and
(\ref{a2}) and the IR mass scale for the particles $\mbar$. Here,
we propose
\beqa
\big[-\pi\mu\gamma(1+b^2/2)\big]^2\big[\frac{\pi\kappa}{2}\gamma(\frac{1+b^2}{2})\big]^2=\left[\frac{{\overline
m}\Gamma(1+\frac{B}{H_2})\Gamma(\frac{1-B}{H_2})}{2^{1+2/H_2}\Gamma(1/H_2)}\right]^{2H_2(1+b^2)}\label{masskap}
\eeqa
for the QFT (\ref{a1}) and
\beqa
\big[-\pi{\tilde\mu}2^{-4b^2-2}\gamma(1+2b^2)\big]\big[\frac{\pi{\tilde\kappa}}{2}\gamma(\frac{1+b^2}{2})\big]^2=
\left[\frac{{\overline m}\Gamma(1+\frac{B}{{\tilde
H}_2})\Gamma(\frac{1-B}{{\tilde H}_2})}{2^{1+2/{\tilde
H}_2}\Gamma(1/{\tilde H}_2)}\right]^{2{\tilde H}_2(1+b^2)}
\label{masskapt} \eeqa
for the QFT (\ref{a2}). One should mention that under the
weak-strong coupling duality transformation $b\leftrightarrow
1/b$, these relations exchange perfectly if the parameters in
action (\ref{a1}) and (\ref{a2}) satisfy the duality relations
\beqa
\pi\mu\gamma(b^2/2)=\Big[\pi{\tilde\mu}\gamma(2/b^2)\Big]^{b^2/2}\qquad
\mbox{and} \qquad \Big[\frac{\pi\kappa b^2
}{2}\gamma\big(\frac{1+b^2}{2b^2}\big)\Big]^{b}=\Big[\frac{\pi{\tilde\kappa}}{2b^2}\gamma(\frac{1+b^2}{2})\Big]^{1/b}
\label{dual1}\ .\eeqa

Using the first relation (\ref{masskap}) in (14) for $n=2$, it is
now possible to compare the expansion (\ref{ceff}) for $n=2$ to
(\ref{ceffTBA}) obtained numerically using (\ref{spect}), (23) and
(\ref{f}). The good agreement\,\footnote{I am very grateful to P.
Dorey and R. Tateo for these numerical checks.} supports the
approach based on the reflection amplitudes, the exact
$\mbar$-$\kappa$-$\mu$ (and their dual) relations given above, the
bulk free energies (\ref{f}) of (\ref{a1}) (and its dual for
(\ref{a2})) as well as the $S$-matrix elements conjectured in
\cite{del}. Due to the following relations
\beqa c^{(2)}_{eff}(R)|_b= {\tilde c}^{(2)}_{eff}(R)|_{1/b} \qquad
\mbox{as}\qquad \ell|_b={\tilde\ell}|_{1/b}\ ,\label{dualcond}
\qquad\eeqa
the weak-strong coupling duality property between the models with
action (\ref{a1}) and (\ref{a2}) for $n=2$ proposed in \cite{del}
is indeed confirmed at the {\it on-shell} level.

\section{Vacuum expectation values of local fields}
 For generic values of the parameter
$n\neq 1,2$, there is no reason to expect the QFTs (\ref{a1}) and
(\ref{a2}) to be integrable. However, for $n=1$ the ShG and BD
integrable models are recovered. For $n=2$ the parafermionic
current $\psi$ is a Majorana fermion. As the scaling limit of the
Ising model with zero external magnetic field is described by a
free Majorana fermion field theory, both models describe different
interactions between a Liouville field theory and a critical Ising
model\,\footnote{Notice that the form of the interaction remains
the same but the definition of the Liouville coupling constant
changes.}. In this case, it can be shown that both models are also
integrable \cite{watts}, and that respective conserved quantities
exchange under weak-strong coupling duality \cite{moda}.

In the previous subsection we considered the UV asymptotic of the
effective central charges in these QFTs. The calculations were
based on the reflection amplitudes, i.e. using CFT data. However,
these functions play also a crucial role in the calculation of
vacuum expectation values in perturbed CFT \cite{onep}. Here, at
the {\it off-shell} level, we will see that VEVs also satisfy such
duality property.

\subsection{Expectation values of primary fields and the
Ising-Liouville dual models}
Here, we consider the expectation values of the simplest local
fields which belong to the Liouville field theory, i.e.
\beqa G_{\tau}^{(n)}(a) = \langle
e^{a\varphi}(x)\rangle_{\tau,n}\qquad \mbox{with}\qquad \tau=1,2\
.\label{Ga}\eeqa
From the discussion in the previous subsection and using once
again CPT framework with (\ref{foncn}), the last term in action
(\ref{a1}) or (\ref{a2}) can be considered as a perturbation. It
is then expected that for any $\tau$ and $n$ the VEV (\ref{Ga})
satisfy the reflection relation
\beqa G^{(n)}_{\tau}(-a)=R_b^{(n)}(-a)\ G^{(n)}_{\tau}(-2Q'+a)\
.\label{reflecn}\eeqa
Let us first study the QFT (\ref{a1}). Instead of the previous
picture,  it is also possible to consider this QFT as a perturbed
Liouville field theory with parameter and coupling ($\mu$,$b$).
The stress-energy tensor for the conformal invariant part is then
\beqa T_{L}(z)=-\frac{1}{2}(\partial\varphi)^2 +
Q\partial^2\varphi \qquad \mbox{with} \qquad
Q=\frac{b}{2}+\frac{1}{b}\ .\label{stresst} \eeqa
Using the CPT framework, the VEV is then expected to satisfy the
reflection relation
\beqa G^{(n)}_{\tau=1}(a)=R_b^{(1)}(-a)|_{\kappa\rightarrow -\mu}\
G^{(n)}_{\tau=1}(2Q-a)\label{reflec1}\eeqa
Obviously, it is not possible to find a solution to these
reflection equations without any strong analyticity assumptions.
Assuming
 that the VEV $G^{(n)}_{\tau=1}(a)$ is a meromorphic function
in $a$, the ``minimal'' solution of the reflection relations
(\ref{reflecn}) and (\ref{reflec1}) can be explicitly obtained
with the result
\beqa
G^{(n)}_{\tau=1}(a)&=&\Big[\big[-\pi\mu\gamma(1+b^2/2)\big]^2\big[\frac{\pi\kappa}{n}\gamma(1/n+b^2/2)\big]^2
\Big]^{\frac{-a^2+2Qa}{2H_n(1+b^2)}}\big[-\pi\mu\gamma(1+b^2/2)\big]^{-\frac{a}{b}}\nonumber
\\
&&
\times\exp{\int_{0}^{\infty}\frac{dt}{t}\Big[a^2e^{-2t}-\frac{\sinh(abt)\Psi_n(t,b,a)}{\sinh(t)
\sinh(tb^2)\sinh(ntb^2/2)\sinh(H_n(1+b^2)t)) }\Big]}\nonumber
\eeqa
where we define for $-(\frac{b^2}{2}+\frac{1}{n})<{\cal
R}e(ab)<\frac{b^2}{2}+1\ $ and
\beqa
\Psi_n(t,b,a)&=&2\sinh((ab+2Q'b)t)\sinh(Qbt)\sinh(ntb^2/2)\cosh(tb^2/2)\nonumber
\\
 &&\qquad +\sinh((ab-2Qb)t)\sinh(nQ'bt)\sinh(tb^2)\ . \label{Gn1}\eeqa
For several other QFTs (Toda, deformed Toda, ...) for which exact
VEVs were proposed explicitly, it has been shown that the
expectation values of the fundamental field \\
$\langle\varphi\rangle$ for $b\rightarrow 0 $ agrees with the same
quantity calculated within perturbation theory. It has also been
shown that semiclassical analysis supports the conjectures for the
VEVs. Consequently, here we expect the same feature to be
satisfied.

We can proceed similarly for the QFT (\ref{a2}). This QFT can be
understood as a perturbed Liouville field theory with parameter
and coupling $({\tilde\mu},2b)$. One has to consider now the
stress-energy tensor with a form similar to (\ref{stresst}) but
with the substitution
\beqa Q \ \longrightarrow \ {\tilde Q}=b+\frac{1}{2b}\ .
\label{charge2}\eeqa
The reflection relation satisfied by the VEV immediately follows
\beqa G^{(n)}_{\tau=2}(a)=R_{2b}^{(1)}(-a)|_{\kappa\rightarrow
-{\tilde\mu}}\ G^{(n)}_{\tau=2}(2{\tilde Q}-a)\ .
\label{reflec1t}\eeqa
As before, using the reflection relations (\ref{reflecn}) and
(\ref{reflec1t}) simultaneously and assuming similarly strong
analytical assumptions we obtain the following conjecture
\beqa
G^{(n)}_{\tau=2}(a)&\!=\!&\Big[\big[\!\!-\pi2^{-4b^2-2}{\tilde\mu}\gamma(1+2b^2)\big]\big[\frac{\pi{\tilde\kappa}}{n}\gamma(1/n+b^2/2)\big]^2
\Big]^{\frac{-a^2+2{\tilde Q}a}{2{\tilde
H}_n(1+b^2)}}\!\big[\!\!-\pi2^{-4b^2-2}{\tilde\mu}\gamma(1+2b^2)\big]^{-\frac{a}{2b}}\nonumber
\\
&&
\times\exp{\int_{0}^{\infty}\frac{dt}{t}\Big[a^2e^{-2t}-\frac{\sinh(abt){\tilde\Psi}_n(t,b,a)}{\sinh(t)
\sinh(tb^2)\sinh(ntb^2/2)\sinh({\tilde H}_n(1+b^2)t))
}\Big]}\nonumber \eeqa
with
\beqa {\tilde\Psi}_n(t,b,a)&=&2\sinh((ab+2Q'b)t)\sinh({\tilde
Q}bt)\sinh(ntb^2/2)\cosh(t/2)\nonumber
\\
 &&\qquad +\sinh((ab-2{\tilde Q}b)t)\sinh(nQ'bt)\sinh(tb^2)\ .\label{Gn2} \eeqa
Here, the integral is convergent if
$-(\frac{b^2}{2}+\frac{1}{n})<{\cal R}e(ab)< b^2+\frac{1}{2}\ .$

For $n=1$, it is straightforward to check that (\ref{Gn1}) and
(\ref{Gn2}) agree respectively with BD and ShG results. For $n=2$,
using the duality transformations of the parameters (\ref{dual1})
it is also easy to check that
\beqa G^{(2)}_{\tau=1}(a)|_{b}=G^{(2)}_{\tau=2}(a)|_{1/b}\ . \eeqa
Finally, using the VEVs (\ref{Gn1}) (or similarly (\ref{Gn2})) for
$n=2$, the bulk free energy of (\ref{a1}) can be calculated as we
have the relation
\beqa \mu G^{(2)}_{\tau=1}(b)=\frac{2(1-B)}{H_2}f^{(2)}_{\tau=1}\
. \eeqa
Using (\ref{masskap}), it is easy to show that the result for
$f^{(2)}_{\tau=1}$ which follows is in perfect agreement with the
expression proposed in (\ref{f}) (and similarly for the QFT
(\ref{a2})).

\subsection{Expectation values of the first descendant fields in the
Ising-Liouville dual models}
By adding the correct counterterms, for $n=2$ both  models
(\ref{a1}) and (\ref{a2}) can now be understood as two different
perturbations of the $N=1$ superLiouville field theory. For
instance, for $\tau=1$ it is given by the action
\beqa {\cal A}_{SL}=\int
d^2x\Big[\frac{1}{2\pi}({\psi}{\overline\partial}\psi +
{\overline\psi}\partial{\overline\psi}) +
\frac{1}{8\pi}(\partial_\nu\varphi)^2 -\kappa\psi{\overline
\psi}e^{-b\varphi} + {\tilde\mu} e^{-2b\varphi}\Big]\ \label{aSL}\
\eeqa
and similarly with the change $\kappa\rightarrow {\tilde\kappa}$
for $\tau=2$. Superconformal transformations in the SL theory are
generated by the super stress-energy tensor ${\hat
T}(z)=-S(z)/2+\theta T(z)$ and similarly for the antiholomorphic
part with $(\theta,{\overline \theta})$ the corresponding
Grassmann coordinates. Here $T(z)$ denotes the usual (bosonic)
holomorphic stress-energy tensor given in eq. (\ref{Tn}) for $n=2$
and $S(z)$ is a spin $3/2$ conserved current:
\beqa S(z) = \psi\partial\varphi + 2Q'\partial \psi
\label{super}\eeqa
and similarly for the antiholomorphic part. With the background
charge given in eq. (\ref{Tn}) for $n=2$, the central charge of
the superLiouville model is then $c_{SL}=\frac{3}{2}(1+8Q'^2)$.

The basics fields in the SL theory are the operators $\sigma_j
\exp(a\varphi)$ which belong either to the Neveu-Schwarz sector
(NS) (for $j=0$) or the Ramond sector (R) (for $j=1$, also called
``twisted'' fields). These primary fields have conformal dimension
\beqa \Delta_{j}(a)=\frac{j(2-j)}{16}-\frac{a(2Q'+a)}{2}\ \ \
\mbox{for}\ \ j=0,1\ . \nonumber\eeqa
Then, using the Laurent expansion
\beqa T(z)=\sum_{n\in{\mathbb Z}}L_n z^{-n-2} \ \ \ \ \ \ \
 \mbox{and} \ \ \ \ \ \ \ \ \ S(z)=\sum_{r\in{\mathbb Z}/2}{S}_r
{z}^{-r-3/2} \ \label{laurent} \eeqa
and similarly for the antiholomorphic part, all the other fields
can be obtained via the action of the Neveu-Schwarz (Ramond)
algebra generators $L_n$, $S_r$ for $n\in{\mathbb Z}$,
$r\in{\mathbb Z}+1/2$ ($n\in{\mathbb Z}$, $r\in{\mathbb Z}$). In
general, we denote these descendent fields
\beqa L_{[n]}{\overline L}_{[m]}S_{[r]}{\overline S}_{[s]} \si_j
e^{a\varphi}\ \equiv \ L_{-n_1}...L_{-n_N}{\overline
L}_{-m_1}...{\overline L}_{-m_K} S_{-r_1}...S_{-r_{N'}}{\overline
S}_{-s_1}...{\overline S}_{-s_{K'}}\si_j
e^{a\varphi}\label{LGnmrs} \eeqa
where $[u]=[-u_1,...,-u_P]$ \ are arbitrary strings. The
descendent fields (\ref{LGnmrs}) and the ones obtained after the
reflection $a\rightarrow -(2Q'+a)$ possess the same quantum
numbers. It is then possible to show that the reflection property
extends to all these descendants and arguments based on CPT
approach suggest the following reflection relation for $j=0$:
\beqa \langle L_{[n]}{\overline L}_{[m]}S_{[r]}{\overline S}_{[s]}
e^{a\varphi}\rangle_{\tau}\ =\ R_{b}^{(2)}(a)\langle L_{[n]}
{\overline L}_{[m]}S_{[r]}{\overline S}_{[s]}
e^{-(2Q'+a)\varphi}\rangle_{\tau}\ \ \ \mbox{and} \ \  \tau=1,2\
,\nonumber \eeqa
where $R_{b}^{(2)}(a)$ is the SL reflection amplitude in the NS
 sector calculated in \cite{StanRash}. For simplicity, here we
only consider the VEV of the simplest descendent fields. Using the
relations (\ref{super}) and (\ref{laurent}) above, we have
\beqa S_{-1/2}{\overline S}_{-1/2} e^{a\varphi} = -a^2 \psib\psi
e^{a\varphi}\ . \eeqa
Consequently, expectation values of these operators in the
perturbed theories (\ref{a1}) or (\ref{a2}) for $n=2$ are expected
to satisfy similar reflection equations. Let us consider the ratio
\beqa H_{\tau}(a)=\frac{\langle\psib\psi
e^{a\varphi}\rangle_{\tau}}{\langle e^{a\varphi}\rangle_{\tau}}\
.\label{Ha}\eeqa
The model with action (\ref{a1}) and $n=2$ can either be
considered as a perturbed SL theory or a perturbed Liouville
theory. As before, approach based on CPT then suggests the
following reflection equations
\beqa H_{\tau=1}(a)\ &=&\ \frac{(2Q'+a)^2}{a^2} \
H_{\tau=1}(-2Q'-a)\ \ \ \ \mbox{for}\ \ \
a\neq 0\ ,\nonumber \\
H_{\tau=1}(a)\ &=&\ \ H_{\tau=1}(2Q-a)\nonumber\ .\eeqa
The ``minimal'' solution of these reflection equations is defined
up to an overall constant. To fix it, it is sufficient to notice
that $H_{\tau=1}(-b)$ can be related with the bulk free energy of
the system as follow
\beqa H_{\tau=1}(-b)\langle
e^{-b\varphi}\rangle_{\tau=1}=\frac{\partial
f_{\tau=1}^{(2)}}{\partial\kappa}\ ,\eeqa
which leads to the result
\beqa \langle\psib\psi e^{a\varphi}\rangle_{\tau=1} =
-\frac{\mbar\
b^2}{2(3+2b^2)^2}\frac{\Gamma(\frac{1}{3+2b^2})\Gamma(\frac{b^2}{3+2b^2})}{\Gamma(\frac{2+b^2}{3+2b^2})}
\gamma\big(\frac{b^2+2-ab}{3+2b^2}\big)\gamma\big(\frac{ab}{3+2b^2}\big)\
\ G^{(2)}_{\tau=1}(a)\ \label{ppbexp}\eeqa
for $a\neq 0$. It is straightforward to do the same analysis for
the model with action (\ref{a2}) for $n=2$. It can be shown that
the result for $\langle\psib\psi e^{a\varphi}\rangle_{\tau=2}$
follows from (\ref{ppbexp}) with the substitution $b\rightarrow
1/b$. In Appendix A, we give a further support to (\ref{ppbexp}).
In conclusion, all previous results for $n=2$ confirm at the {\it
off-shell} level the weak-strong coupling duality.

\section{Application to integrable coupled minimal models and the homogeneous sine-Gordon $SU(3)_2$ model}
For $n=2$, the QFT with action (\ref{a2}) possesses interesting
applications. As is well known, the ``minimal model'' ${\cal
M}_{p/p'}$ with central charge $c=1-6\frac{(p-p')^2}{pp'}$ can be
obtained from the Liouville model. Consequently, adding the
counterterm \ ${\tilde \mu}e^{-2b\varphi}$\  in action (\ref{a2}),
the resulting QFT can be identified with a minimal model
interacting with a critical Ising model if we substitute
\beqa b\rightarrow i\beta,\ \ \ \ \ \ {\tilde \mu}\rightarrow
-{\tilde\mu}\ , \ \ \ \ \ \ {\tilde \kappa}\rightarrow
i{\tilde\kappa}\ , \label{sub}
 \eeqa
and make the choice either
\beqa
 \beta^2=\beta^2_+=p/2p' \ \ \ \ \ \ \ \mbox{or}\ \ \ \ \ \ \
\beta^2=\beta^2_-=p'/2p \ \ \ \ \ \ \ \ \mbox{with} \ \ \ p<p'\
\label{conbet}. \eeqa
For both values of the coupling and using the background charge
(\ref{charge2}) which ensures the local conformal invariance of
the Liouville field theory, the conformal dimension of the
perturbing term becomes\ $\Delta(\psi{\overline
\psi}e^{-i\beta\varphi})=3\beta^2/2$, which is relevant for
$\beta^2<2/3$\ .

In the following, we define $\{\Phi_{rs}\}$ as the set of primary
fields with conformal dimensions
\beqa \Delta_{rs}=\frac{(p'r-ps)^2-(p-p')^2}{4pp'} \ \ \ \
\mbox{for}\ \ \ 1\leq r<p,\ \ 1\leq s < p' \ \ \mbox{and}\ \ \
p<p'.\label{dim} \eeqa
Using the Coulomb gas representation, they are simply related with
the vertex operators of the Liouville field theory through the
relation
\beqa \Phi_{rs}(x)=N^{-1}_{rs}\exp(i\eta^{rs}\varphi(x))\ \ \ \ \
\ \mbox{with}\ \ \ \ \ \ \ \
\eta^{rs}=-\frac{(1-r)}{2\beta}+(1-s)\beta\ ,\label{primaire}
\eeqa
where we have introduced the normalization factors $N_{rs}$. These
numerical factors depend on the normalization of the primary
fields. Here, they are chosen in such a way that the primary
fields satisfy the conformal normalization condition:
\beqa \langle\Phi_{rs}(x)\Phi_{rs}(y)\rangle_{CFT} \ =\
\frac{1}{|x-y|^{4\Delta_{rs}}} \ .\label{norm} \eeqa
For further convenience, we write these coefficients
$N_{rs}=N(\eta^{rs})$ where:
\beqa
N(\eta)=\big[-\pi{\tilde\mu}\gamma(-2\beta^2)\big]^{-\frac{\eta}{2\beta}}
\big[\frac{\Gamma(2\beta^2-2\eta\beta)\Gamma(1/2\beta^2+\eta/\beta)
\Gamma(2-2\beta^2)\Gamma(2-1/2\beta^2)}{\Gamma(2-2\beta^2+2\eta\beta)
\Gamma(2-1/2\beta^2-\eta/\beta)\Gamma(2\beta^2)\Gamma(1/2\beta^2)}\Big]^{\frac{
1}{2}} \ .\label{normalisation}
\eeqa

Taking specific values of the coupling constant in action
(\ref{a2}), it is now possible to obtain non-perturbative
information for two planar systems which interact through a
relevant operator preserving integrability. The first system -
described by the free Majorana fermion part - is identified with a
critical Ising model denoted ${\cal{M}}_{3/4}$ using the standard
conventions. The second one - obtained from the QG restriction of
the Liouville field theory - is identified with a minimal model
denoted ${\cal{M}}_{p/p'}$. Then, the action can be written either
\beqa {\cal{A}}\ &=& \ {\cal{M}}_{3/4}\ +\ {\cal{M}}_{p/p'}\ +
{\lambda} \int d^2x\ \epsilon(x)\Phi_{12}(x)\qquad
\mbox{for}\qquad
\beta^2=\beta^2_+\label{action12}\\
\mbox{or}\qquad  {\hat{\cal{A}}}\ &=& \ {\cal{M}}_{3/4}\ + \
{\cal{M}}_{p/p'}\ + {\hat\lambda} \int d^2x\
\epsilon(x)\Phi_{21}(x)\qquad \mbox{for}\qquad \beta^2=\beta^2_-\
,\label{action21} \eeqa
where the parameters \ $\lambda$ and $\hat\lambda$ \ characterize
the strength of the interaction and the energy operator of the
Ising model is defined by $\epsilon\equiv i{\overline\psi}\psi$.
Also, both QFTs make sense if $3p<4p'$ in (\ref{action12}) and
$3p'<4p$ in (\ref{action21}), which ensure the perturbation to be
relevant.

For imaginary values of the coupling $b=i\beta$, it is expected that the QFT (\ref{a2})
 possesses complex soliton solutions which interpolate between the
degenerate vacua. It can be shown that this model possesses the QG
symmetry associated with $U_q(B(0,1)^{(1)})$ where $q$ is the
deformation parameter. Naturally, there are good reasons to
believe that the $S$-matrix of this model can be expressed in
terms of the $R$-matrix associated with this deformed affine Lie
superalgebra. Then, in the following we {\it assume} that a
breather-particle
 identification holds by comparing the resulting $S$-matrix elements of the
lowest-breathers (breathers with lowest mass) with the $S$-matrix
elements  for the quantum particles in the real coupling case.
Denoting $M$ as the mass of the lowest kink, we suggest the
identification
\beqa \mbar = 2M\sin\big(\frac{\pi\xi}{2-\xi}\big)\ \ \ \ \ \ \ \
\mbox{with}\ \ \ \ \ \ \ \
 \ \xi=\frac{\beta^2}{1-\beta^2}\label{rel}\ .
\eeqa
In the following, we will study successively the models with
action (\ref{action12}) and (\ref{action21}). However, the vacuum
structure in both cases is not clearly understood so the prefactor
associated with the QG restriction is ommited for simplicity.
Consequently, we denote all expectation values\
$\langle...\rangle\equiv\langle0|...|0\rangle$\ where $|0\rangle$
is one of the many ground states.

Let us first consider the coupled minimal models with action (\ref{action12}).
With the previous identification and using the exact $\mbar$-${\tilde\kappa}$-${\tilde\mu}$
relations (\ref{masskapt}) proposed in the previous section,
it is now straightforward to obtain the following exact $M$-$\lambda$ relations
\beqa \la^2=\frac{1}{\pi^2}
\frac{\gamma(\frac{3\xi-1}{1+\xi})\gamma(\frac{1-\xi}{1+\xi})}
{\gamma^2(\frac{1}{2(1+\xi)})} \left[ \frac{\pi
M\Gamma(\frac{1+\xi}{2-\xi})}
{\Gamma(\frac{\xi}{2-\xi})\Gamma(\frac{1}{2-\xi})}
\right]^{\frac{2(2-\xi)}{1+\xi}} \ \ \ \ \ \  \ \mbox{for}\ \ \ \
\ \ \xi=\frac{p}{2p'-p}\ .\label{Mlambda} \eeqa
Consequently, according to (\ref{Mlambda}) and $\beta^2<2/3$, the
coupled minimal models (\ref{action12}) develop a massive spectrum
for:
\beqa (i) \ \ \ &&  1/3<\xi<1, \ \ \ \ \  {\mathfrak I}m(\lambda)
= 0 \ \ \ \
\mbox{i.e.}\ \ \ \frac{1}{2} < \frac{p}{p'} < 1\ ,\label{condietii}\\
(ii) \ \ \ && 0<\xi<1/3, \ \ \ \ \ {\mathfrak R}e(\lambda) = 0 \ \
\ \ \ \mbox{i.e.}\ \ \ 0 < \frac{p}{p'} <\frac{1}{2}\ .\nonumber
\eeqa
In particular, the condition $(i)$ is always satisfied for the
coupled {\it unitary} minimal  models defined by (\ref{action12})
with $p'=p+1$ . Notice that for $p=2$, $p'=3$ the model
(\ref{action12}) corresponds to an off-critical Ising model as we
have $\Phi_{12}={\mathbb I}$ in this case. If we make the
identification $\mbar=2M\sin(\pi/3)\equiv M_{SG}$ where $M_{SG}$
is the SG soliton mass, then the $M$-$\lambda$ relation given
above\,\footnote{The parameter $\lambda$ is real and we obtain
$M_{SG}=2\pi|\la|$ .} coincides with the one associated with the
off-critical Ising model, as expected.

Although the model (\ref{a2}) with imaginary coupling is very
different from the real coupling case in its physical content (the
model contains solitons and excited solitons), there are good
reasons to believe that the expectation values obtained in the
real coupling case provide also the expectation values for
imaginary coupling. Similarly to the analysis done for the ShG and
BD models \cite{onep}, we obtain the VEV of primary operators
which belong to the second minimal model using eqs. (\ref{Gn2})
for $b=i\beta_+$, (\ref{primaire}) and (\ref{normalisation})
\beqa \langle\Phi_{rs}(x)\rangle =
\left[\frac{\pi^2\lambda^2\gamma^2(\frac{1}{2(1+\xi)})(1+\xi)^{\frac{4-2\xi}{1+\xi}}}
{2^{\frac{4}{1+\xi}}\gamma(\frac{3\xi-1}{1+\xi})\gamma(\frac{1-\xi}{1+\xi})}
 \right]^{\frac{(1+\xi)}{2-\xi}\Delta_{rs}}
\! \exp{\cal Q}_{\epsilon 12}((1+\xi)r-2\xi s)\ .\label{VEV12}
\eeqa
The function  \ ${\cal Q}_{\epsilon 12}(\th)$ \ for\ \
$|\th|<2\xi$\ \ and \ $\frac{1}{3}<\xi<1$ \ is given by the
integral
\beqa {\cal Q}_{\epsilon 12}(\th)= \int_0^{\infty} \frac{dt}{t}
\Big(\frac{\Psi_{\epsilon
12}(\th,t)}{\sinh((1+\xi)t)\sinh(t\xi)\sinh((2-\xi)t)}
-2\Delta_{rs}e^{-2t} \Big)\nonumber \eeqa
with
\beqa
\Psi_{\epsilon 12}(\th,t)&=& \frac{\sinh(t)}{2}\Big[\cosh(\th t)-\cosh((1-\xi)t)\Big] + \Big[\cosh(\th t)\cosh((2-\xi)t)\nonumber\\
&&\ -\
\sinh((2-\xi)t)\sinh((1-\xi)t)-\cosh(t)\Big]\sinh(\frac{\big(1-\xi)t}{2}\big)\cosh\big(\frac{(1+\xi)t}{2}\big)\
\nonumber \eeqa
and defined by analytic continuation outside this domain. Notice
that eq. (\ref{VEV12}) satisfies
\beqa \langle\Phi_{rs}(x)\rangle= \langle\Phi_{p-r\
p'-s}(x)\rangle\ . \eeqa
Finally, the expectation value (\ref{VEV12}) can be used to derive
the bulk free energy \ \ $f_{\epsilon 12} = -\limi{V\rightarrow
\infty}\frac{1}{V} \ln Z$\ \ where $V$ is the volume of the 2D
space and $Z$ is the singular part of the partition function
associated with action (\ref{action12}). The result for the bulk
free energy follows from the analytic continuation of (\ref{f})
and eq. (\ref{rel}), i.e.
\beqa f_{\epsilon
12}=-\frac{M^2}{2}\frac{\sin(\frac{\pi\xi}{2-\xi})\sin(\frac{\pi}{2-\xi})}{\sin
(\frac{\pi(1+\xi)}{2-\xi})}\label{f12}\ . \eeqa
If we look at the action (\ref{action12}), we can now use the
relation \ \ $\partial_\lambda f_{\epsilon 12}\ =\
\langle\epsilon\Phi_{12}\rangle$\ \ to deduce the expectation
value
\beqa \langle\epsilon(x)\Phi_{12}(x)\rangle =
-\frac{1}{\lambda}\left[\frac{\lambda^2\pi^2
\gamma^2(\frac{1}{2(1+\xi)})}{\gamma(\frac{3\xi-1}{1+\xi})\gamma(\frac{1-\xi}{1+\xi})}\right]^\frac{1+\xi}{2-\xi}
\frac{1+\xi}{2-\xi}\frac{\Gamma^2(\frac{1}{2-\xi})}{\Gamma^2(\frac{2(1-\xi)}{2-\xi})\Gamma^2(\frac{1+\xi}{2-\xi})}
\frac{\sin(\frac{\pi}{2-\xi})}{\sin(\frac{\pi\xi}{2-\xi})\sin(\frac{\pi(1+\xi)}{2-\xi})}\
. \label{vevpert12}\eeqa

Instead of taking $\Phi_{12}$ inside the expectation value written
above, one can now consider the one associated with more general
primary operator of the second minimal model. The corresponding
VEV follows from the dual result of eq. (\ref{ppbexp}), i.e. we
obtain
\beqa
\langle\epsilon(x)\Phi_{rs}(x)\rangle&=&-\left[\frac{\lambda^2\pi^2
\gamma^2(\frac{1}{2(1+\xi)})}{\gamma(\frac{3\xi-1}{1+\xi})\gamma(\frac{1-\xi}{1+\xi})}\right]^\frac{1+\xi}{2(2-\xi)}
\frac{1+\xi}{2-\xi}\gamma(\frac{1}{2-\xi})\frac{\gamma(\frac{1-\xi+(1+\xi)r-2\xi
s }{4-2\xi})}{\gamma(\frac{3-\xi+(1+\xi)r-2\xi s}{4-2\xi})}\
\langle\Phi_{rs}(x)\rangle\ . \eeqa

Let us now turn to the coupled minimal models with action
(\ref{action21}). In this case, the condition $4p>3p'$ guarantees
that the perturbing operator is relevant. Then, the vacuum
structure is expected to be similar to that of (\ref{action12}).
Using the substitutions
\beqa p\leftrightarrow p', \ \ \ \ \ r\leftrightarrow s,\ \ \ \ \
\xi \longrightarrow
 \frac{1+\xi}{3\xi-1}\label{analyticcont}
\eeqa
in the previous expressions, the breather-particle relation is now
given by
\beqa {\overline m}=2M\sin(\frac{\pi(1+\xi)}{5\xi-3})\label{reldu}
\eeqa
and the exact relation between the mass of the lightest kink $M$
and $\hat\lambda$ follows
\beqa {\hat\la}^2=\frac{1}{\pi^2}
\frac{\gamma(\frac{1}{\xi})\gamma(\frac{\xi-1}{2\xi})}
{\gamma^2(\frac{3\xi-1}{8\xi})} \left[ \frac{\pi M
\Gamma(\frac{4\xi}{5\xi-3})}{\Gamma(\frac{3\xi-1}{5\xi-3})\Gamma(\frac{1+\xi}{5\xi-3})}
\right]^{\frac{5\xi-3}{2\xi}}\ . \label{masshatla}\eeqa
Then, for the coupled minimal models defined by (\ref{action21}),
the massive phase corresponds to the domain:
\beqa (iii) \ \ \ \frac{3}{5}\ < \xi <1, \ \ \ \ \ {\mathfrak
I}m({\hat\lambda})=0\ \ \ \mbox{i.e.} \ \ \ \frac{3}{4}\ <\
\frac{p}{p'}\ <\ 1\ . \eeqa
From eq. (\ref{VEV12}) and (\ref{analyticcont}), we obtain the
following expression for the VEV
\beqa \langle\Phi_{rs}(x)\rangle =
\left[\frac{\pi^2{\hat\lambda}^2\gamma^2(\frac{3\xi-1}{8\xi})(2\xi)^{\frac{5\xi-3}{2\xi}}}
{2^{\frac{1+\xi}{2\xi}}\gamma(\frac{1}{\xi})\gamma(\frac{\xi-1}{2\xi})}\right]^{\frac{4\xi}{5\xi-3}\Delta_{rs}}
\! \exp{\cal Q}_{\epsilon 21}((1+\xi)r-2\xi s)\ .\label{VEV21}
\eeqa
The function  \ ${\cal Q}_{\epsilon 21}(\th)$ \ for \ \ $|\th| <
2\xi$\ \ is given by the integral
\beqa {\cal Q}_{\epsilon 21}(\th)= \int_0^{\infty} \frac{dt}{t}
\Big(\frac{\Psi_{\epsilon
21}(\th,t)}{\sinh((1+\xi)t)\sinh(4t\xi)\sinh((5\xi-3)t)}
-2\Delta_{rs}e^{-2t} \Big)\nonumber \eeqa
with
\beqa \Psi_{\epsilon 21}(\th,t)\!\!&=&\!\!
\frac{\sinh((3\xi-1)t)}{2}\Big[\cosh(2\th t)-\cosh((2-2\xi)t)\Big]
+ \Big[\cosh(2\th t)\cosh((5\xi-3)t) \nonumber\\
&-&\!\sinh((2-2\xi)t)\sinh((5\xi-3)t)-\cosh((3\xi-1)t)\Big]\sinh((\xi-1)t)\cosh((2\xi)t)\
\nonumber\eeqa
and is defined by analytic continuation outside this domain. As
was considered above, the exact bulk free energy can be also
calculated from the results of the previous section,
\beqa f_{\epsilon
21}=-\frac{M^2}{2}\frac{\sin(\frac{\pi(3\xi-1)}{5\xi-3})
\sin(\frac{\pi(1+\xi)}{5\xi-3})}{\sin(\frac{\pi(4\xi)}{5\xi-3})}\label{f21}\
. \eeqa
Finally, this latter expression provides us the VEV
\beqa
\langle\epsilon(x)\Phi_{21}(x)\rangle=-\frac{1}{\hat\lambda}\left[\frac{{\hat\lambda}^2\pi^2
\gamma^2(\frac{3\xi-1}{8\xi})}{\gamma(\frac{1}{\xi})\gamma(\frac{\xi-1}{2\xi})}\right]^\frac{4\xi}{5\xi-3}
\frac{4\xi}{5\xi-3}\frac{\Gamma^2(\frac{3\xi-1}{5\xi-3})}{\Gamma^2(\frac{4\xi-4)}{5\xi-3})\Gamma^2(\frac{4\xi}{5\xi-3})}
\frac{\sin(\frac{\pi(3\xi-1)}{5\xi-3})}{\sin(\frac{\pi(1+\xi)}{5\xi-3})\sin(\frac{\pi(4\xi)}{5\xi-3})}\
,\label{vevpert21} \eeqa
whereas for a more general primary operator of the second minimal
model, we now obtain
\beqa
\langle\epsilon(x)\Phi_{rs}(x)\rangle&=&-\left[\frac{{\hat\lambda}^2\pi^2
\gamma^2(\frac{3\xi-1}{8\xi})}{\gamma(\frac{1}{\xi})\gamma(\frac{\xi-1}{2\xi})}\right]^\frac{2\xi}{5\xi-3}
\frac{4\xi}{5\xi-3}\gamma(\frac{3\xi-1}{5\xi-3})\frac{\gamma(\frac{\xi-1-(1+\xi)r+2\xi
s}{5\xi-3})}{\gamma(\frac{4\xi-2-(1+\xi)r+2\xi s}{5\xi-3})}\
\langle\Phi_{rs}(x)\rangle\ . \eeqa

Some checks of these expressions are desirable. For instance, the
case $p=5$ and $p'=6$ in action (\ref{action21}) corresponds to a
critical Ising model coupled to a critical $\mathbb{Z}_{3}$-Potts
model through the interaction $\epsilon {\cal E}$ where ${\cal
E}\equiv \Phi_{21}$ is the leading energy operator of the
$\mathbb{Z}_3$-Potts model. It is rather interesting to recall
that the decoupled critical Ising-3-state Potts models appear in
the phase diagram of the ${\mathbb Z}_6$ spin model in the
vicinity of a renormalization group fixed point. Using continuum
field theory approach, which is valid in the scaling region around
this point, Zamolodchikov's counting argument \cite{counti} can be
used to show that the $\epsilon {\cal E}$ perturbation preserves
conserved charges of spin $\pm 3$, $\pm 5$. As was suggested in
\cite{doreyt} and confirmed here by explicitly constructing the
QFT, this perturbation of the Ising-3-state Potts model is
therefore integrable. Although the exact $S$-matrix for these
coupled models is not known explicitly, a TBA system based on
$E_7$ Dynkin diagram has already been considered as a good
candidate \cite{Rav}. The exact relation (\ref{masshatla}) as well
as the breather-particle relation can been checked using this TBA
system. The good agreement\,\footnote{P. Dorey and R. Tateo,
private communication} supports the exact results above.

Let us now explain how the above results can be relevant in the
study of the $SU(3)_2$-HSG  model. The simplest HSG  model is the
complex sine-Gordon model associated with an integrable
perturbation of the WZNW-coset model $SU(2)_k/U(1)$, whereas more
complicated HSG theories can be viewed as interacting copies of
complex sine-Gordon theories. At classical level, the
corresponding equations of motion correspond to non-abelian Toda
theories which are known to be integrable. At quantum level it has
also been shown that integrability is preserved \cite{mir0}, and
assuming factorization of scattering process $S$-matrices have
been proposed \cite{mir}. Among the generalizations of the complex
sine-Gordon model, the $SU(3)_2$-HSG can describe a WZNW-coset
model $SU(3)_2/(U(1))^2$ with central charge $c=6/5$ perturbed by
an operator with conformal dimension $\Delta_{pert}=3/5$, as was
shown in \cite{mir0}. On the other hand, the case $p=4$ and $p'=5$
in action (\ref{action12}) leads to a critical Ising model coupled
to a tricritical Ising model (with a total central charge
$c=1/2+7/10=6/5$ in the UV limit). It is then interesting to
notice that conformal dimensions\
$\Delta(\epsilon(x)\Phi_{12}(x))=\Delta(\Phi_{13}(x))=3/5$.
Obviously, in the deep UV the operator content of the coupled
models (\ref{action12}) is bigger than the operator content of the
$SU(3)_2$-HSG, which only consists of operators of conformal
dimension $0$, $1/2$, $1/10$ and $3/5$ \cite{hsg}. However, the
operator content of (\ref{action12}) possesses a closed subset of
operators $\{1,\epsilon,\Phi_{12},\epsilon\Phi_{12},\Phi_{13}\}$
with conformal dimensions given above. Consequently, using the
notations of \cite{hsg} for the $SU(3)_2$-HSG, we propose the
identification ${\cal O}_{0,2}^{0,1}\equiv \Phi_{12}$, the trace
of the energy momentum tensor
$\Theta\in\{\epsilon\Phi_{12}\cup\Phi_{13}\}$ whereas the
remaining operator is the Ising energy density $\epsilon$. Then,
the action associated with the $SU(3)_2$-HSG model can be seen as
a subsector of the following (integrable \cite{Phil}) action:
\beqa S_{HSG[SU(3)_2]}\ \thicksim \ {\cal{M}}_{3/4}\ +\
{\cal{M}}_{4/5}\ + \lambda \int d^2x \Phi_{pert}(x) \eeqa
with \ $\Phi_{pert}(x) = \epsilon(x)\Phi_{12}(x) +
\rho\Phi_{13}(x)$ and $\rho$ is a $c-$number. At the special value
$\rho=0$, one recovers (\ref{action12}). From the above analysis
(\ref{condietii}), it follows that the $SU(3)_2$-HSG is a massive
theory in agreement with the results of \cite{mir0}. Also, taking
the value $\xi=2/3$ in (\ref{rel}) gives $\mbar=2M$, i.e. the
formation of stable particles via fusing of the soliton mass $M$
is not possible in agreement with \cite{mir2}. At $\rho=0$, its
bulk free energy follows from (\ref{f12}) whereas the exact
relation between the soliton mass $M$ and the parameter in the
Lagrangian $\lambda$ is given in (\ref{Mlambda}) for $\xi=2/3$.

Accepting the conjectures (\ref{VEV12}) and (\ref{VEV21}), one can
then make interesting predictions for numerical values of VEVs. We
report the reader to the Appendix B where various examples are
considered.

\section{Relation to the coupled Ising-XY model}
Over the years, the critical behavior of the two dimensional
Ising-XY model, consisting of Ising and XY models coupled through
their energy densities, has been studied numerically in some
detail. A Hamiltonian for this model has been proposed which
writes
\beqa {\cal H}/kT=
-\sum_{i,j}\big[(A+B\si_i\si_j)\cos(\th_i-\th_j) + C \si_i\si_j
\big]\label{ffxy}\eeqa
where $A$, $B$ and $C$ are effective couplings. The model with
$A\neq B$ is relevant for the anisotropic frustrated XY model
\cite{gra} and anti-ferromagnetic RSOS model \cite{nijs}  whereas
the subspace $A=B$ is relevant for the isotropic FFXY model or its
one dimensional quantum version \cite{gra2}. In this latter case,
a phase diagram has been proposed (see figure 1). It consists of
three branches joining at a point $P$ in the ferromagnetic region
$A>0$, $A+C>0$. One of these branches corresponds to a single
transition with simultaneous loss of Ising and XY order and the
other two to separate Kosterlitz-Thouless (KT) and Ising
transitions. Monte Carlo transfer-matrix methods \cite{MC} yields
that the central charge seems to vary continuously from
$c\thickapprox 1.5$ near $P$ to $c\thickapprox 2$ at $T$, which
contradicts the hypothesis that the line $PT$ can be simply
described in terms of a superposition of critical Ising and XY
models with $c=1.5$ as was suggested by Foda in \cite{foda}. The
only possibility would be the existence of a parameter changing
along the line $PT$ that does not affect the symmetry. As was
argued in \cite{kost} there are three possible explanations: $(a)$
The system is not conformally invariant; $(b)$ The result is an
artifact of limited strip widths; $(c)$ It is a new effect.
Consequently, due to the limited strip widths of the numerical
analysis, there are some reasons to believe that the phase diagram
analysis is not yet complete, in particular, first order
transitions may appear along the line denoted $PT$. More recent
numerical analysis of the Ising-XY model based on the coupled
Ising-RSOS model \cite{Lee} also supports this hypothesis. Then,
if we opt for scenario $(a)$, it is natural to consider all
possible integrable perturbations of a superposition of critical
Ising and XY model as a starting point to study the vicinity of
the point $P$. It can be shown that there exists only three kinds
of integrable perturbations which can provide interesting
candidates\,\footnote{Notice that the continuum limit of the
generalized Coulomb-gas representation of the FFXY model
containing fractional charges is nothing but the (conformally
invariant) action of a free Majorana fermion and a free boson.}:
the supersymmetric sine-Gordon model \cite{susg} which has been
suggested by Foda or the models with action (\ref{a1}) and
(\ref{a2}) for $n=2$ and imaginary coupling.

\vspace{5mm}

\centerline{\epsfig{file=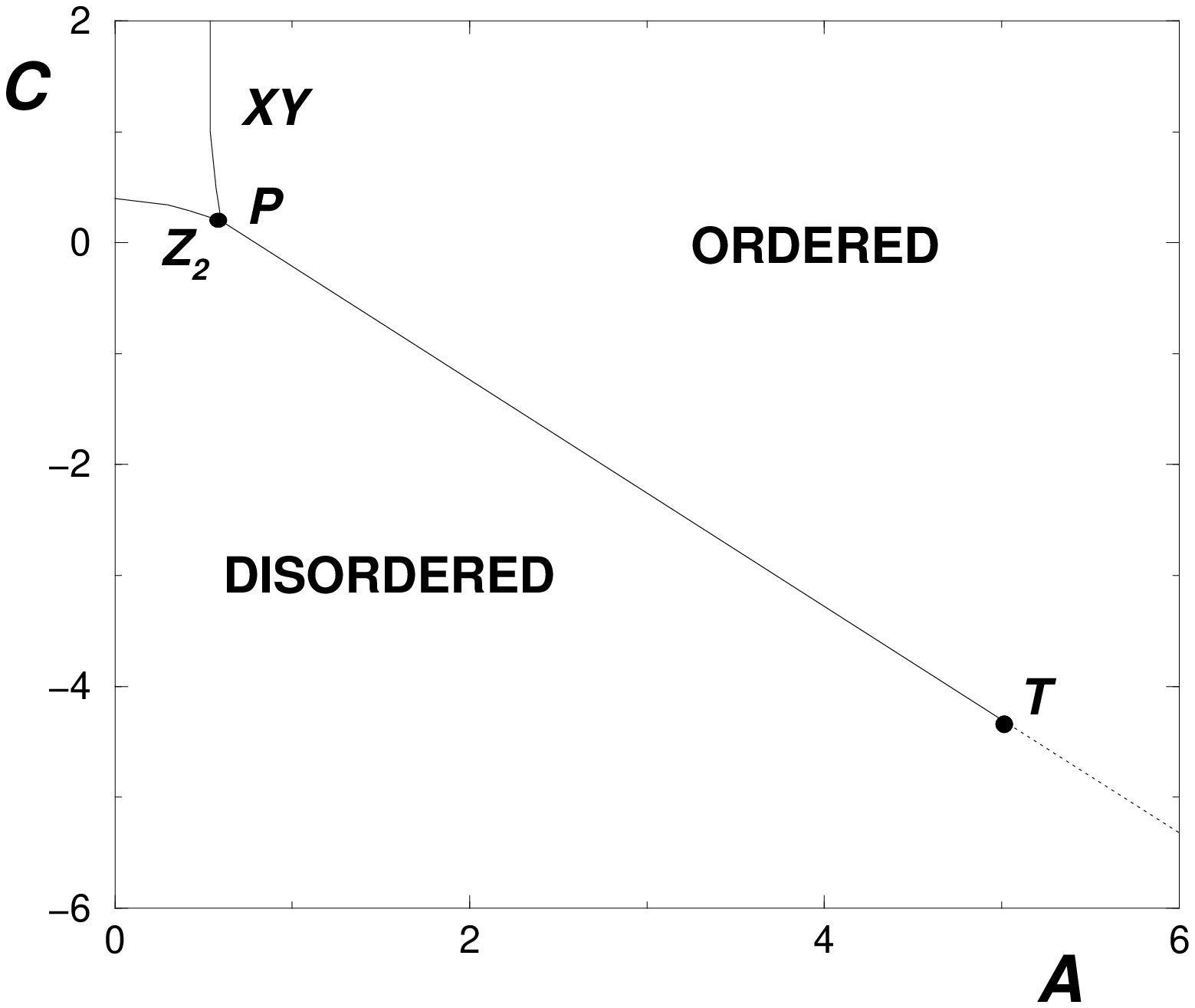,height=50mm,width=60mm}}
\vspace{5mm}
\begin{center}
{\small \underline{Figure 1}: The phase diagram for the coupled
Ising-XY models}
\end{center}

\vspace{0.3cm}

Consequently, it would be rather interesting to obtain
non-perturbative results for these cases. As some exact off-shell
results (VEVS, bulk free energy,...) for the supersymmetric SG
model can be found in \cite{basfat}, here we naturally focus on
action (\ref{a2}) for $n=2$\,\footnote{We have seen in the
previous section that the model (\ref{a2}) for $n=2$ can describe
a critical RSOS model coupled to a critical Ising model through
their energy densities. Furthermore, since for $n=2$ both models
with actions (\ref{a1}) and (\ref{a2}) possess the same limit at
$\beta^2=1/2$, it is sufficient to focus on one only.}. Using the
analytic continuation $b\rightarrow i\beta$ and ${\tilde \mu}
\rightarrow -{\tilde \mu}$, we will consider in the vicinity of
the point $P$ of the phase diagram the following action:
\beqa {\cal A}_{\beta}=\int
d^2x\Big[\frac{1}{2\pi}({\psi}{\overline\partial}\psi +
{\overline\psi}\partial{\overline\psi}) +
\frac{1}{8\pi}(\partial_\nu\varphi)^2
-{\tilde\kappa}\psi{\overline \psi}e^{-i\beta\varphi} -
2{\tilde\mu}\cos(2\beta\varphi)\Big]\ .\label{aXYI}\ \eeqa
Before going further, let us recall known results for the XY model
- which also corresponds to the nonlinear $O(2)$ $\sigma$-model.
Kosterlitz and Thouless showed that spin configurations are
mixture of topologically trivial configurations (called spin
waves) and a gas of vortices with integer topological charge. Both
are decoupled and the vortices interact through a logarithmic
potential which is therefore identical to a two dimensional
Coulomb gas. At a specific finite critical coupling, it has been
shown rigorously \cite{fro} that this Coulomb gas possesses a
phase transition. In the vicinity of that point, only vortices of
topological charges $\pm 1$ are important and higher charge
vortices can be neglected. In the Coulomb gas formalism, various
configurations are then associated with the operators ${\cal
O}_{e,m}$ with dimension $d_{e,m}=e^2/R^2+mR^2/4$ and spin
$s_{e,m}=em$. In particular, the ``electric'' $e$ and ``magnetic''
$m$ charges exchange each other under the weak-strong coupling
duality $R\leftrightarrow 2/R$. Also, the ``electric'' operators
can be written in terms of vertex operators in the following way
\beqa {\cal O}_{e,0}(x)={\cal N}^{-1}(e\beta)\exp(ie\varphi/R)(x)
\ \ \ \ \mbox{with}\ \ \ e\in{\mathbb Z} \eeqa
where we choose the normalization factor (the mass scale $M$ has
been introduce in (\ref{rel}) )
\beqa {\cal N}(\eta)=\left[\Big(\frac{\pi M
\Gamma(\frac{1}{2-3\beta^2})}{2^{\frac{2(1-\beta^2)}{2-3\beta^2}}\Gamma(\frac{1-\beta^2}{2-3\beta^2})
\Gamma(\frac{\beta^2}{2-3\beta^2})}\Big)^{2\beta^2-1}
\big(\pi{\tilde\mu}2^{4\beta^2-2}\gamma(1-2\beta^2)\big)^{1/2}\right]^{-\frac{\eta}{\beta}}\
. \eeqa

Returning to action (\ref{aXYI}) it is now quite natural to take
$\beta=1/{\sqrt 2}$ in order to consider the Ising-XY model. For
this value and $R={\sqrt 2}$, the operators ${\cal O}_{\pm 2,0}$
become marginal as well as the last part of the action (cosine
term) (\ref{aXYI}). For $e>2$, operators are irrelevant. Then, we
obtain the following action
\beqa {{\cal{A}}_{1/{\sqrt 2}}} &=& {\cal M}_{3/4} +
\frac{1}{8\pi}\int d^2x (\partial_\nu\varphi)^2 +
\Lambda_{\beta=1/{\sqrt 2}} \int d^2x\ \epsilon(x){\cal
O}_{-1,0}(x)\ .\label{actionisingXY}\eeqa
For general values of $\beta$, the exact relation between the mass
scale $M$ and the parameter $\Lambda_\beta$ follows from section 2
and eq. (\ref{rel}):
\beqa
\Lambda_{\beta}=\frac{2}{\pi\gamma(\frac{1-\beta^2}{2})}\left[\frac{\pi
M\Gamma(\frac{1}{1-\beta^2})}{2^{\frac{2(1-\beta^2)}{2-3\beta^2}}\Gamma(\frac{1-\beta^2}{2-3\beta^2})
\Gamma(\frac{\beta^2}{2-3\beta^2})}\right]^{1-\beta^2}\ . \eeqa
Also, using the conventions defined above, an exact expression for
the VEV of the ``electric'' operators can be obtained:
\beqa \langle{\cal O}_{e,0}(x)\rangle&=&\left[\frac{\pi M
\Gamma(\frac{1}{2-3\beta^2})}
{2^{\frac{2(1-\beta^2)}{2-3\beta^2}}\Gamma(\frac{\beta^2}{2-3\beta^2})\Gamma(\frac{1-\beta^2}{2-3\beta^2})}\right]^{e^2\beta^2}\nonumber
\\
&&\ \ \ \ \times  \ \exp\int_0^\infty\frac{dt}{t}\Big[-e^2\beta^2
e^{-2t}-\frac{\sinh(e\beta^2t)\Psi_{\beta}(t)}{\sinh(t)\sinh(t\beta^2)
\sinh((2-3\beta^2)t)}\Big]\nonumber \eeqa
with
\beqa
\Psi_{\beta}(t)&=&2\sinh((1-(e+1)\beta^2)t)\sinh((1/2-\beta^2)t)\cosh(t/2)\nonumber
\\
&& -\sinh((1+(e-2)\beta^2)t)\sinh((1-\beta^2)t)\ . \eeqa

For the choice $\beta=1/{\sqrt 2}$, the model
(\ref{actionisingXY}) becomes massive if \ ${\cal
I}m(\Lambda_{\beta=1/{\sqrt 2}})=0$, corresponding to a first
order phase transition. One can then check that $\langle{\cal
O}_{\pm 1,0}\rangle=\langle{\cal O}_{0,\pm 1}\rangle$ and
$\langle{\cal O}_{\pm 2,0}\rangle=0$ as expected. As the cosine
term in the action is marginal,  one can show that exactly the
same results can be obtained if we had started from action
(\ref{a1}) instead. This is not surprising as both models possess
the same limit at $\beta^2=1/2$.

\section{Neveu-Schwarz sector of perturbed $N=1$ supersymmetric unitary minimal models}
Instead of considering a restriction of action (\ref{a2}), it is
also interesting to study action (\ref{a1}) for $n=2$ at specific
imaginary values of the coupling $b$. As it is known, the minimal
series of superconformal {\it unitary} models can be described
from the superLiouville field theory using the analytic
continuation $b\rightarrow i\beta$ in (\ref{aSL}) and taking the
specific value
\beqa \beta^2=\frac{K}{K+2}\ \ \ \ \ \ \ \ \ \mbox{with}\ \ \  \ \
\ \ \ \ \ K\geq 2 \ .\nonumber\eeqa
Their corresponding central charge is \ \ $
c_{SUSY}=\frac{3}{2}\big(1-\frac{8}{K(K+2)}\big)$\ \ and the
finite number of primary operators belonging to the NS sector are
labelled by the conformal dimensions (for $r-s\in {2\mathbb Z}$)
\beqa \Delta^{(NS)}_{rs}=\frac{(r(K+2)-s K)^2-4}{8K(K+2)}\ \ \ \ \
\mbox{with}\ \ \ \ 1\leq r <K \ ,\ \ 1 \leq s < K+2\ .\nonumber
\eeqa
Using the vertex operator representation, they can be written in
terms of the exponential fields of the superLiouville theory as
follows:
\beqa \Phi^{(NS)}_{rs}(x)={\cal
N}^{-1}_{rs}\exp(i\eta^{rs}\varphi(x))\ \ \ \mbox{with} \ \ \
\eta^{rs}=\frac{(1-r)}{2\beta}-\frac{(1-s)}{2}\beta \ \ \
\mbox{and}  \ \ r-s\in {2\mathbb Z}\label{primairesusy} \eeqa
where we have introduced the normalization factors ${\cal
N}_{rs}$. Choosing a condition similar to (\ref{norm}), they can
be expressed in terms of the reflection amplitude in the NS sector
of the superLiouville field theory \cite{StanRash} as \ ${\cal
N}_{rs}\equiv {\cal N}^{(NS)}(\eta_{rs})$\ where
\beqa {\cal
N}^{(NS)}(\eta)&=&\big[\frac{\pi{\kappa}}{2\beta^2}\gamma\big(\frac{1-\beta^2}{2}\big)\big]^{\frac{\eta}{\beta}}\nonumber
\\
&\times&\Big[\frac{\Gamma(1/2+\beta^2/2+\eta\beta)\Gamma(1/2+1/2\beta^2-\eta/\beta)
\Gamma(3/2-\beta^2/2)\Gamma(3/2-1/2\beta^2)}{\Gamma(3/2-\beta^2/2-\eta\beta)
\Gamma(3/2-1/2\beta^2+\eta/\beta)\Gamma(1/2+\beta^2/2)\Gamma(1/2+1/2\beta^2)}\Big]^{\frac{
1}{2}} \ .\label{normalisationsusy} \nonumber\eeqa

Adding the perturbing term with conformal dimension \
$\Delta_{pert}=\Delta_{SL}(e^{b\varphi})=-b^2-1/2$\ in
(\ref{aSL}), the analytic continuation $b \rightarrow i\beta$ then
gives perturbed $N=1$ supersymmetric minimal model\,\footnote{To
describe other perturbations of supersymmetric minimal models, one
may have also considered the analytic continuation of the model
with action (\ref{a2}). However, for $\beta^2=K/(K+2)$ the
perturbation corresponds to $\Phi_{15}^{(NS)}$, which is relevant
only for $K=4$. On the other hand, the analytic continuation of
action (\ref{a1}) at the (dual) value $\beta^2=(K+2)/2$ with
$K\leq 4$ leads to a relevant perturbation, namely
$\Phi^{(NS)}_{31}$. For simplicity we will not consider this case
here.}. Notice that for any values of $K$, the perturbation is
relevant and is identified to $\Phi_{13}^{(NS)}$ with conformal
dimension\ $\Delta^{(NS)}_{13}=\frac{K-2}{2(K+2)}$\ . The
resulting action writes
\beqa {\tilde{\cal{A}}} &=& {\cal{M}}^{N=1}_{K} + {\tilde\lambda}
\int d^2x\ \Phi^{(NS)}_{13}(x)\ .\label{actionsusy} \eeqa
The exact relation between the parameter ${\tilde \lambda}$ and
the mass scale $\mbar$ introduced in eq. (\ref{spect}) can be
obtained using ${\tilde\lambda}=-\mu{\cal N}^{(NS)}(\beta)$ with
the result
\beqa {\tilde\lambda}=-\frac{(1+\txi)}{\pi
\gamma(\frac{2+\txi}{2+2\txi})}\Big[\frac{\Gamma(\frac{1+4\txi}{2+2\txi})\Gamma(\frac{3+2\txi}{2+2\txi})}
{(2\txi-1)\Gamma(\frac{3}{2+2\txi})\Gamma(\frac{1+2\txi}{2+2\txi})}\Big]^{\frac{1}{2}}
\Big[\frac{\mbar\Gamma(\frac{3}{3+\txi})\Gamma(\frac{1+\txi}{3+\txi})}
{2^{\frac{4}{3+\txi}}\Gamma(\frac{1}{3+\txi})}\Big]^{\frac{3+\txi}{1+\txi}}\
. \label{lambdamsusy}\eeqa
For any values of $K$ we have $1/2<\beta^2<1$. Consequently, the
model (\ref{actionsusy}) develops a massive phase for ${\mathfrak
I}m({\tilde\lambda}) = 0$\ . The calculation of the bulk free
energy leads to the result
\beqa
f^{SUSY}_{13}=-\frac{\mbar^2}{8}\frac{\sin(\frac{\pi}{3+{\tilde\xi}})}
{\sin(\frac{\pi(1+{\tilde\xi})}{3+{\tilde\xi}})\sin(\frac{\pi{\tilde\xi}}{3+{\tilde\xi}})}\label{fsusy}\eeqa
where we have $\tilde\xi=K/2$, which gives the following
expression for the exact vacuum expectation value of the
perturbing operator
\beqa
\langle\!\Phi^{(NS)}_{13}\!\rangle&\!\!=\!\!&-\frac{1}{\tilde\lambda}
\Big[\pi^2{\tilde\lambda}^2\gamma^2\big(\frac{\txi+2}{2+2\txi}\big)
\frac{\Gamma(\frac{3}{2+2\txi})\Gamma(\frac{1+2\txi}{2+2\txi})}{\Gamma(\frac{1+4\txi}{2+2\txi})\Gamma(\frac{3+2\txi}
{2+2\txi})}\frac{2\txi-1}{4(1+\txi)^2}\Big]^{\frac{1+\txi}{3+\txi}}\frac{1+\txi}{3+\txi}\nonumber
\\
&&\ \ \ \ \ \ \ \  \ \times\ \
\frac{2^{\frac{4}{3+\txi}}\Gamma^2(\frac{1}{3+\txi})}
{\Gamma^2(\frac{3}{3+\txi})\Gamma^2(\frac{1+\txi}{3+\txi})}\frac{\sin(\frac{\pi}{3+\txi})}
{\sin(\frac{\pi(1+\txi)}{3+\txi})\sin(\frac{\pi\txi}{3+\txi})}\
.\nonumber \eeqa
From the exact expression of the vacuum expectation value
(\ref{Gn1}) and the definition of the normalization factor, for
more general VEVs of primary fields one obtains:
\beqa \langle\Phi^{(NS)}_{rs}\rangle &=&
\Big[\frac{\mbar\Gamma(\frac{3}{3+{\tilde\xi}})\Gamma(\frac{1+{\tilde\xi}}{3+{\tilde\xi}})(1+{\tilde\xi})}{2^{1+\frac{2}{3+{\tilde\xi}}}
\Gamma(\frac{1}{3+{\tilde\xi}})}\Big]^{2\Delta_{rs}} \nonumber \\
 &&\times \exp\int_{0}^{\infty}\!\frac{dt}{t}\Big[-2\Delta_{rs}
e^{-4t} - \frac{{\cal F}((1+{\tilde\xi})r-{\tilde\xi}
s,{\tilde\xi},t)}{2\sinh(2(1+{\tilde\xi})t)\sinh(2{\tilde\xi}
t)\sinh((6+2{\tilde\xi})t)}\Big]\ ,\nonumber
 \eeqa
where we introduce the function
\beqa {\cal
F}(\th,{\tilde\xi},t)&=&\sinh((1-\th)t)\big(\cosh((\th+3+2\xi)t)+\cosh((\th+3)t)-\cosh((\th-1)t)\nonumber
\\ && -\cosh((\th-1-2\xi)t)+(\cosh((\th-3-2\xi)t)-\cosh((\th-7-2\xi)t)\nonumber \\
&& +\cosh((\th+9+2\xi)t)-\cosh((\th+5+2\xi)t))/2\big)\ .\eeqa

For $K=2$, i.e. $\txi=1$ the central charge of the supersymmetric
minimal model is $c_{SUSY}=0$. Then the perturbing operator
reduces to $\Phi^{(NS)}_{13}\equiv {\mathbb I}$, i.e. the identity
operator. In this case, one can check that the above relation
$\tilde\lambda$-$\mbar$ gives $\tilde\lambda=-\mbar^2/8$ which
implies $f^{SUSY}_{13}=\tilde\lambda$, as expected.

For $K=3$, the model with action (\ref{actionsusy}) describes an
integrable perturbation of the tricritical Ising model. On the
other hand, the same model can be obtained starting from an
integrable perturbation of a {\it nonsupersymmetric} minimal
models, as ${\cal M}_{4/5}\equiv{\cal M}_3^{N=1}$. Using notations
of the previous section, we have the correspondence
$\Phi_{12}\equiv\Phi^{(NS)}_{13}$. For this case, the VEVs, exact
relations between the mass scale of the particles and the
parameter in the action have been proposed in \cite{onep}. It is
straightforward to check that the above results perfectly agree
with these ones as long as the mass scale $\mbar\equiv M$ of
\cite{onep}.

For $K=6$, the central charge of the conformal part becomes
$c=5/4$, i.e. the resulting model ${\cal M}_{6}^{N=1}$ is
identified with a critical ${\mathbb Z}_6$ spin model perturbed by
the thermal operator with conformal dimension \ $D_1=1/4$ \
denoted $\epsilon^{(1)}$ in \cite{fatzam}. In general, ${\mathbb
Z}_n$ spin models are the natural generalizations of the Ising
model, when the spin variable takes its values in group ${\mathbb
Z}_n$ \cite{fatzam}. These self-dual models possess critical
points which are associated with ${\mathbb Z}_n$ parafermionic
CFTs with central charge $c=\frac{2(n-1)}{n+2}$ \cite{fatzam}
briefly described in beginning of section 2. Besides the
parafermionic symmetry, these CFTs possess ${\cal W}(A_{n-1})$
symmetry and can be described by the ${\cal M}_{n+1}(A_{n-1})$
minimal model at the specific coupling $\beta^2=\frac{n+1}{n+2}$.
It has been shown that integrability is preserved by adding the
perturbing first thermal operator $\epsilon^{(1)}$ with conformal
dimension $2/(n+2)$. This operator is anti self-dual, i.e. his
sign changes under duality transformation. Depending of the sign
of the parameter characterizing the strength of the perturbation,
the perturbed theory will be either in ordered or disordered
phase. For general values of $n$, exact relation between this
parameter and the mass of the particles have been proposed in
\cite{norma}. However, to compare our results to the ones
associated with the perturbed model ${\cal M}_{7}(A_{5})$, we need
to identify the lowest kink in both models. Similarly to the case
of the coupled models in section 4, for general values of $\tilde
\xi$ here we expect the identification
\beqa \mbar = 2M \sin\big(\frac{\pi{\tilde \xi}}{3+{\tilde
\xi}}\big)\eeqa
where $M$ denotes the mass of the lowest kink. Taking ${\tilde
\xi}=3$ and using the substitution above in (\ref{lambdamsusy})
and (\ref{fsusy}), it is straightforward to check that previous
results coincide perfectly with the ones in \cite{norma}.

\section{Concluding remarks}
In a first part, we have studied in detail two of the simplest
dual representations of integrable deformations of affine Toda
theories, which corresponds to a critical Ising model coupled with
a Liouville field theory depending on the coupling $b$. In the
deep UV, the effective central charges which characterize the
behavior of both models are obtained explicitly. Also, we propose
the exact relations between the parameters in the Lagrangians (UV
data) and the mass scale of the particles (IR data), which are
necessary to perform properly the TBA analysis. Various vacuum
expectation values, namely the one associated with the primary
field $\langle e^{a\varphi}(x)\rangle$ and the first descendant
field $\langle \psib\psi e^{a\varphi}(x)\rangle$, as well as the
bulk free energy are obtained explicitly. All previous results are
shown to exchange under a weak-strong coupling duality
$b\leftrightarrow 1/b$, confirming the duality relation between
both Lagrangian representations conjectured in \cite{del,moda}.

In the second part, we consider various applications of these
results. From the model (\ref{a2}) for $n=2$, we were able to
describe several kind of coupled models. First, we studied in
detail the case $b^2=-p/2p'$ or $b^2=-p'/2p$ from which we obtain
the coupled minimal models (\ref{action12}) or (\ref{action21}),
respectively. The main difference between both actions is the
presence of an extra term in (\ref{action12}), associated with the
$\Phi_{13}$ primary operator. It comes from the counterterm in the
real coupling QFT. In both cases, we propose exact relation
between the mass of the lowest breather $\mbar$ and the mass of
the lowest kink $M$, as well as the exact relation between the
parameters $\lambda$ or ${\hat \lambda}$ and $M$. For a critical
Ising model coupled with {\it unitary} minimal models, the QFTs
(\ref{action12}) and (\ref{action21}) are found to be massive.
Exact VEVs and bulk free energies are obtained. Some special cases
have been independently checked and shown to agree using different
methods$^{8}$. Among these coupled models, we propose an
identification between a special case ($\rho=0$) of the
$SU(3)_2$-HSG model for a finite resonance parameter and a
subsector of the Ising-Tricritical Ising models coupled through
energy-energy. In particular, we find that the lowest breather
disappears from the spectrum, unstable as expected \cite{mir2}.

Secondly, we propose to study the vicinity of the critical point
$P$ in the phase diagram of the coupled Ising-XY model depicted in
figure 1. Three candidates, integrable perturbations of a $c=3/2$
CFT, can be considered. The supersymmetric sine-Gordon suggested
in \cite{foda} or the models (\ref{a1}) or (\ref{a2}) for
imaginary coupling and $n=2$. Here we focused on the QFT
(\ref{actionisingXY}) obtained from (\ref{a2}) for $n=2$ as the
other results can be found in \cite{basfat}. The model is a
massive QFT, corresponding to a first order phase transition along
the line $PT$ in figure 1. Exact VEV of the electric/magnetic
operator $\langle {\cal O}_{\pm 1,0}\rangle=\langle {\cal
O}_{0,\pm 1 }\rangle$ is calculated.

From the model (\ref{a1}) for $n=2$, we study in detail the NS
sector of the QG restriction associated with $b^2=-K/K+2$, which
leads to the $\Phi_{13}^{SUSY}$ integrable perturbation of $N=1$
unitary minimal models. We show that results agree with the known
ones obtained for the perturbed Tricritical Ising model
\cite{onep} for $K=3$ and the perturbed ${\mathbb Z}_6$
parafermionic CFTs \cite{norma} for $K=6$. Finally we would like
to add a
few remarks:\\
\\
$\bullet$ {\bf Scattering theory of solitons for imaginary
coupling.}\\
For imaginary coupling both models studied here admit a quantum
group symmetry based on ${\cal U}_{q}(A^{(4)}(0,2))$ and ${\cal
U}_{q}(B^{(1)}(0,1))$, respectively. Then, it would be interesting
to construct the corresponding $R$-matrices in such a way to
obtain the exact $S$-matrices for the scattering of solitons. The
identification of the scattering amplitudes for the lowest
breathers with the $S$-matrix for quantum particles in the real
coupling case should confirm eq. (\ref{rel}). Also, the restricted
$R$-matrix with respect to the different subalgebras should
provide the scattering amplitudes in the coupled models described
here. Understanding the vacuum structure of all the models
described above would also allow one to fix the form of the
prefactor associated with the QG restriction.\\
\\
$\bullet$ {\bf Homogeneous sine-Gordon $SU(3)_2$ model.}\\
As we mentioned above, the perturbing field
$\Phi_{pert}\backsimeq\Theta(x)$ with conformal dimension $3/5$
can be written as \ $\Phi_{pert}(x) = \epsilon(x)\Phi_{12}(x) +
\rho\Phi_{13}(x)$ where $\rho$ is a $c-$number. For a finite
resonance parameter \cite{mir0,hsg} the $SU(3)_2$-HSG model
contains only two self-conjugate solitons with masses $M$ and
$M'$. Then, it would be rather interesting to obtain the exact
relation between the ratio $M/M'$ and the parameter $\rho$. With
the $\lambda$-$M$ relation ($M$ is the lightest mass of the two
self-conjugate solitons) obtained from eq. (\ref{Mlambda}) for
$\xi=2/3$, i.e. $|\lambda|=0.2790872531...M^{4/5}$ and using
(\ref{VEV12}) and (\ref{vevpert12}) it would be straightforward to
obtain the expectation values $\langle{\cal
O}_{0,2}^{0,1}(x)\rangle$ and $\langle\Phi_{pert}(x)\rangle$ in
the HSG-$SU(3)_2$ model for a finite resonance parameter.

Furthermore, it is well-known that any correlation function of
local fields\ $\langle{\cal O}(r){\cal O }(0)\rangle$\ in the
short-distance limit can be reduced down to one-point functions \
$\langle{\cal O}'(r)\rangle$ by successive application of the
operator product expansion \cite{2,3}. Along the line of
\cite{3,mag}, using the exact results obtained here as well as the
three-point functions calculated in \cite{StanRash} and the
relation between the HSG-$SU(3)_2$ and the coupled
Ising-tricritical Ising models proposed in section 4, the UV
behavior of the two-point correlation functions can be studied. On
the other hand, the long-distance behavior (IR) of the two-point
correlation functions in the HSG-$SU(3)_2$ model has been studied
in detail in \cite{hsg} using the so-called form-factors approach.
Then, it would be interesting to compare both results which should
agree in a good approximation in
the intermediate region $Mr\sim 1$.\\
\\
$\bullet$ {\bf Ramond sector of $N=1$ superconformal unitary
minimal models.}\\
In this work we mainly focused on the NS sector of both models.
However one may be interested in the R sector, for instance, the
VEVs of the primary fields\ $\langle\sigma
e^{a\varphi}\rangle_\tau$. In this case we have to consider the
reflection amplitudes in the R sector of the SL field theory
instead of the NS one. This reflection amplitude has been
calculated in \cite{StanRash} with the result
\beqa
R^{(R)}_{b}(a)=\Big[\frac{\pi\kappa}{2b^2}\gamma\big(\frac{1+b^2}{2}\big)\Big]^{\frac{2(a+Q')}{b}}
\frac{\Gamma(\frac{1}{2}-(a+Q')b)\Gamma(\frac{1}{2}-(a+Q')/b)}{\Gamma(\frac{1}{2}+(a+Q')b)\Gamma(\frac{1}{2}+(a+Q')/b)}\
\label{refR} \nonumber\eeqa
where $2Q'=b+1/b$. If we define
$G_{\tau}^{(R)}(a)\equiv\langle\sigma e^{a\varphi}\rangle_{\tau}$
for the VEVs of the R primary fields in both models, instead of
(\ref{reflecn}) we have
\beqa G^{(R)}_{\tau}(-a)=R_b^{(R)}(-a)\ G^{(R)}_{\tau}(-2Q'+a)\
.\label{reflecR}\nonumber\eeqa
For the QFT (\ref{a1}), the minimal solution of this reflection
equation together with (\ref{reflec1}) can be obtained as before.
The result can be written in terms of the NS one, i.e.
\beqa \langle\sigma e^{a\varphi}\rangle_{\tau=1} =
\langle\sigma\rangle_{\tau=1} G_{\tau=1}^{(2)}(a)\
\exp{\int_{0}^{\infty}\frac{dt}{t}\frac{\sinh(abt)\sinh((ab-2Qb)t)f(t,b)}{\sinh(t)
\sinh(tb^2)\sinh(H_2(1+b^2)t)) }}\nonumber \eeqa
with
\beqa f(t,b) = \sinh((b^2+1)t)-\sinh(b^2t)-\sinh(t) \nonumber
\eeqa
and \ $\langle\sigma\rangle_\tau$ is the VEVs of the Ising spin
field in the QFT (\ref{a1}). The result for the QFT (\ref{a2}) is
straightforward.

Considering specific values of the coupling constant in the QFT
(\ref{a1}) it is then possible to obtain the VEVs $\langle\sigma
\Phi_{rs}\rangle$ in the coupled minimal models studied in sect.
4, which provides the leading term of the two-point function
between
 operators which belong to the critical Ising model and minimal model,
respectively. Also, from the QFT (\ref{a2}) one obtains the
primary operators in the R sector of the perturbed $N=1$
superconformal minimal models.\\

To conclude, as we mentioned briefly in the Introduction the
models studied here belong to a more general family of deformed
Toda models based on Lie superalgebras. In these cases, previous
analysis can be performed along the same line. At specific values
of the coupling, beyond describing different kind of coupled
models, these series are identified with integrable perturbation
of CFTs associated with $WB(0,r)$ algebras with central charge
\beqa c_{WB(0,r)}=(r+1/2)\big(1-2r(2r-1)\frac{(p-p')^2}{pp'}\big)
\nonumber\eeqa
for $p,p'\geq 2r-1$. We intend to return to these models in the
future.\\

\noindent{\bf Acknowledgements:} I am very grateful to P. Dorey
and R. Tateo for useful discussions and several numerical checks
and Al. Zamolodchikov for important comments and reading the
manuscript. I thank Ph. Lecheminant for discussions about
one-dimensional quantum spin systems and bringing my attention to
\cite{hsg}, and especially
 P. Simon for discussions about the Ising-XY model and
interest in this work. I also thank V. Fateev, J. M. Kosterlitz
and R. Sasaki for discussions. Part of this work has been done in
the Department of Mathematics, University of York, UK. Work
supported in part by Marie Curie fellowship HPMF-CT-1999-00094 and
JSPS fellowship.

\vspace{1cm}

\centerline{\bf Appendix A: Normalization of the VEV of the first
descendant field}
\vspace{0.5cm}
To calculate the VEV of the first descendant field \ $\langle \psi
\psib \exp(a\varphi) \rangle_{\tau}$, we considered $H_{\tau}(a)$
defined in eq. (\ref{Ha}) as the minimal solution of certain
reflection equations. However, we used the exact result for the
bulk free energy to fix its overall coefficient. It is then rather
important to possess an independent derivation for this
coefficient, based on the ``resonance condition''\,\footnote{I am
very grateful to Al. Zamolodchikov for suggesting this check.}.

For instance, let us consider the analytic continuation
$b\rightarrow i\beta$ and $\mu\rightarrow -\mu$ of model
(\ref{a1}) with $n=2$. In the free field theory, the composite
field \ $\psi \psib \exp(i\alpha\varphi)$ \ is spinless with scale
dimension
\beqa D\equiv \Delta+{\overline \Delta}=1+\alpha^2.\label{dim}
\eeqa
For generic values of the coupling $\beta$ some divergences arise
in the VEVs of the descendant fields due to the perturbation in
(\ref{a1}) with imaginary coupling. They are generally cancelled
if we add specific counterterms which contain spinless local
fields with cutt-off dependent coefficients. If the perturbation
is relevant, a finite number of lower scale dimension couterterms
are then sufficient. However, this procedure is regularization
scheme dependent, i.e. one can always add finite counterterms. For
generic values of $\alpha$ this ambiguity in the definition of the
renormalized expression for the descendant fields can be
eliminated by fixing their scale dimensions to be (\ref{dim}).
Here, this situation arises if two fields, say ${\cal O}_{\al}$
and ${\cal O}_{\al'}$, satisfy the ``resonance condition''
\cite{des}
\beqa D_\al=D_{\al'} + k(1-\beta^2) + l(2-\beta^2)\ \ \ \
\mbox{with}\ \ \ \ (k,l)\in {\mathbb N} \eeqa
associated with the ambiguity
\beqa {\cal O}_{\al}\longrightarrow{\cal O}_{\al} +
{\kappa}^{k}{\mu}^{l}{\cal O}_{\al}. \eeqa
In this specific case one says that the renormalized field ${\cal
O}_\al$ has an $(k|l)$-th {\it resonance} \cite{des} with the
field ${\cal O}_{\al'}$. In particular, the first descendant field
\ $\psi \psib \exp(i\alpha\varphi)$ has a $(1|0)$ resonance with
the field \ $\exp(i(\alpha-\beta)\varphi)$ at $\alpha=0$. If one
looks at the short distance behavior in $r=|x|\rightarrow 0$ of
the two-point function
\beqa \langle e^{i\alpha_1\varphi}(x) \
\psi\psib(0)e^{i\alpha_2\varphi}(0)\rangle_{\tau=1}\ \ \ \ \ \
\mbox{with}\ \ \al_1+\al_2=\al \ \label{2pt},\eeqa
the contributions brought by $H_{\tau=1}(i\al)$ and $\langle
\exp(-i\beta\varphi)\rangle_{\tau=1}$ have the same power behavior
in $r$. Furthermore, $H_{\tau=1}(i\al)$ and the coefficient in
front of the second VEV in the short distance expansion of
(\ref{2pt}) both exhibit a pole at $\al=0$. By analogy with the
method used in \cite{des}, we require that the divergent
contributions compensate each other. This leads to the relation
\beqa {\cal R}es_{\al=0}{H_{\tau=1}}(i\al) =
\frac{\pi\kappa}{\beta} \ \langle
\exp(-i\beta\varphi)\rangle_{\tau=1}\ .\label{residu}
 \eeqa
It is straightforward to check that (\ref{Ha}) with $\tau=1$ and
$b\rightarrow i\beta$ indeed satisfies this requirement. In
particular, for \ $\tau=1,2$\  this gives a further support to\
$\langle \psi \psib e^{-b\varphi}\rangle_{\tau}$ \ and the exact
bulk free energy proposed for both models.

\vspace{1cm}

\centerline{\bf Appendix B: Numerical values for coupled minimal
models}
\vspace{0.5cm}
$\bullet$ {\bf Two energy-spin coupled Ising models.}\\
It corresponds to $p=3$, $p'=4$ in (\ref{action12}), i.e.
$\xi=3/5$ in (\ref{VEV12}). In this case $\Phi_{12}$ is the spin
operator of the second model in (\ref{action12}) with conformal
dimension $\Delta_{12}=1/16$ whereas $\Phi_{13}$ with conformal
dimension $\Delta=1/2$ is the energy operator. We obtain
\beqa
\langle\Phi_{12}(x)\rangle&=& 1.281110557...M^{1/8};\nonumber\\
\langle\epsilon(x)\Phi_{12}(x)\rangle&=&
7.253910604...M^{9/8}\nonumber \eeqa
where the parameter $\lambda$ is related to the mass of the lowest
kink by\ \ $\lambda^2=0.07660622552...M^{7/4}$.\\

\vspace{0.1cm}

$\bullet$ {\bf Two energy-energy coupled Ising-tricritical Ising
models.}\\
The case $p=4$, $p'=5$ in (\ref{action12}) describes a critical
Ising model which interacts with a tricritical Ising model through
their leading energy density operators. For the second model
$\Phi_{12}$ has conformal dimension $\Delta_{12}=1/10$. It
corresponds to \ $\beta^2=2/5$\ i.e.\ \ $\xi=2/3$\  in
(\ref{VEV12}). This model also contains the sub-leading energy
density operator $\Phi_{13}$ with $\Delta_{13}=3/5$ (``vacancy
operator''), two magnetic operators $\Phi_{22}$ with
$\Delta_{22}=3/80$, $\Phi_{21}$ with $\Delta_{21}=7/16$ and
$\Phi_{14}$. We have for instance
\beqa
\langle\Phi_{12}(x)\rangle&=& 1.495279412...M^{1/5};\nonumber\\
\langle\Phi_{22}(x)\rangle&=& 1.133076821...M^{3/40};\nonumber\\
\langle\epsilon(x)\Phi_{12}(x)\rangle&=&
4.478886063...M^{6/5}\nonumber \eeqa
where the parameter $\lambda$ is related to the mass of the lowest
kink by  \ \ $\lambda^2=0.07788969384...M^{8/5}$.\\

\vspace{0.3cm}

$\bullet$ {\bf Two Ising-$A_5$ RSOS coupled models.}\\
It corresponds to the choice $p=5$, $p'=6$ in (\ref{action12}).
The $A_5$ RSOS model  possesses a primary operator $\Phi_{12}$
with conformal dimension $\Delta_{12}=1/8$ and $\Phi_{22}$ with
$\Delta_{22}=1/40$.  Taking $\xi=5/7$ in (\ref{VEV12}), we obtain
\beqa \langle\Phi_{22}(x)\rangle&=&
1.081616064...M^{1/20};\nonumber\\
\langle\Phi_{12}(x)\rangle&=&
1.673156742...M^{1/4};\nonumber \\
\langle\epsilon(x)\Phi_{12}(x)\rangle&=&
3.478873566...M^{5/4}\nonumber \eeqa
where the parameter $\lambda$ is related to the mass of the lowest
kink by $\lambda^2=0.07848322325...M^{3/2}$.\\

\vspace{0.3cm}

$\bullet$ {\bf Two energy-spin coupled
Ising-tricritical Ising models.}\\
 For $p=4$, $p'=5$ in
(\ref{action21}), the Ising energy operator is coupled to the
subleading spin operator of the tricritical Ising model. For
$\xi=2/3$\  in (\ref{VEV21}) we have for instance \ \
$\langle\Phi_{22}(x)\rangle=  1.2148...M^{3/40}$\ \ where the
parameter $\lambda$ is related to the mass of the lowest kink by
$\lambda^2=0.0319842...M^{1/4}$.
\\

\vspace{0.3cm}

$\bullet$ {\bf Two energy-energy coupled Ising-3-state
Potts models.}\\
The case $p=5$, $p'=6$ in (\ref{action21}) describes a critical
Ising model coupled to a critical 3-state Potts models through
their energy density operator $\epsilon$ and $\Phi_{21}$ with
conformal dimension $\Delta_{21}=2/5$, respectively.  It
corresponds to \ $\xi=5/7$\ in (\ref{VEV21}). The 3-state Potts
model also contains the primary operator $\Phi_{23}$ - the spin
operator - with $\Delta_{23}=1/15$. We obtain for instance\ \
 $\langle\Phi_{23}(x)\rangle=
1.3378...M^{2/15}$\ \ where the parameter $\lambda$ is related to
the mass of the lowest kink by $\lambda^2=0.0420507...M^{2/5}$.

\end{document}